\documentclass[useAMS,usenatbib,usegraphicx]{mn2e}

\usepackage{amsmath}
\usepackage{amssymb}
\usepackage{url}
\usepackage[dvips]{color} 
\usepackage{gensymb}

\newcommand{\Msun}{\ensuremath{\mathrm{M}_\odot}}

\newcommand{\Ni}{\ensuremath{^{56}\mathrm{Ni}}}

\newcommand{\Mej}{\ensuremath{M_\mathrm{ej}}}
\newcommand{\Eej}{\ensuremath{E_\mathrm{ej}}}
\newcommand{\Mni}{\ensuremath{M_\mathrm{Ni}}}

\title[Ultra-stripped SN properties]
{
Light-curve and spectral properties of ultra-stripped core-collapse supernovae
leading to binary neutron stars
}
\author[T. J. Moriya et al.]
{Takashi J. Moriya$^{1,2}$\thanks{takashi.moriya@nao.ac.jp}, 
Paolo A. Mazzali$^{3,4}$,
Nozomu Tominaga$^{5,6}$,
Stephan Hachinger$^{7}$, \newauthor
Sergei I. Blinnikov$^{8,9,6}$,
Thomas M. Tauris$^{10,2}$,
Koh Takahashi$^{11}$,
Masaomi Tanaka$^{1,6}$,\newauthor
Norbert Langer$^{2}$, and
Philipp Podsiadlowski$^{12,2}$
\\
$^{1}$
Division of Theoretical Astronomy, National Astronomical Observatory of Japan, National Institutes of Natural Sciences, \\
2-21-1 Osawa, Mitaka, Tokyo 181-8588, Japan \\
$^{2}$
Argelander Institute for Astronomy, University of Bonn, Auf dem H{\"u}gel
71, D-53121 Bonn, Germany \\
$^{3}$
Astrophysics Research Institute, Liverpool John Moores University,
IC2, Liverpool Science Park, 146 Browlow Hill, Liverpool L3 5RF, UK \\
$^{4}$
Max Planck Institute for Astrophysics, Karl-Schwarzschild-Stra{\ss}e 1,
D-85748 Garching, Germany \\
$^{5}$
Department of Physics, Faculty of Science and Engineering, Konan
University, 8-9-1 Okamoto, Kobe, Hyogo 658-8501, Japan \\
$^{6}$
Kavli Institute for the Physics and Mathematics of the Universe (WPI),
The University of Tokyo Institutes for Advanced Study, \\
The University of Tokyo, 5-1-5 Kashiwanoha, Kashiwa, Chiba 277-8583, Japan \\
$^{7}$
Leibniz Supercomputing Centre (LRZ), Bavarian Academy of Sciences and Humanities, Boltzmannstra{\ss}e 1, D-85748 Garching, Germany \\
$^{8}$
Institute for Theoretical and Experimental Physics, Bolshaya Cheremushkinskaya ulitsa 25, 117218 Moscow, Russia \\
$^{9}$
All-Russia Research Institute of Automatics, Sushchevskaya ulitsa 22, 127055 Moscow, Russia \\
$^{10}$
Max Planck Institute for Radio Astronomy, Auf dem H\"ugel 69, D-53121 Bonn, Germany \\
$^{11}$
Department of Astronomy, Graduate School of Science, The University of Tokyo, Hongo 7-3-1, Bunkyo, Tokyo 113-0033, Japan \\
$^{12}$
Department of Physics, University of Oxford, Denys Wilkinson Building, Keble Road, Oxford OX1 3RH, UK
}

\begin{document}

\date{Accepted 2016 December 8. Received 2016 December 8; in original form 2016 August 25}

\pagerange{\pageref{firstpage}--\pageref{lastpage}} \pubyear{2016}

\maketitle
\label{firstpage}

\begin{abstract}
We investigate light-curve and spectral properties of ultra-stripped
core-collapse supernovae. Ultra-stripped supernovae are the explosions of heavily
stripped massive stars which lost their envelopes via binary interactions with 
a compact companion star. They eject only $\sim 0.1~\Msun$ and may be the main 
way to form double neutron-star systems which eventually merge emitting strong 
gravitational waves. We follow the evolution of an ultra-stripped supernova 
progenitor until iron core collapse and perform explosive nucleosynthesis 
calculations. We then synthesize light curves and spectra of ultra-stripped 
supernovae using the nucleosynthesis results and present their expected 
properties. Ultra-stripped supernovae synthesize
$\sim 0.01~\Msun$ of radioactive \Ni, and their typical peak luminosity is
around $10^{42}~\mathrm{erg~s^{-1}}$ or $-16$~mag. Their typical rise time is
$5-10$~days. Comparing synthesized and observed spectra, we find that
SN~2005ek, some of the so-called calcium-rich gap transients, and SN~2010X may
be related to ultra-stripped supernovae. If these supernovae are actually
ultra-stripped supernovae, their event rate is expected to be about 1 per cent
of core-collapse supernovae. Comparing the double neutron-star merger rate
obtained by future gravitational-wave observations and the ultra-stripped
supernova rate obtained by optical transient surveys identified with our
synthesized light-curve and spectral models, we will be able to judge whether
ultra-stripped supernovae are actually a major contributor to the binary 
neutron star population and provide constraints on binary stellar evolution.
\end{abstract}


\section{Introduction}\label{sec:introduction}

Multiplicity of massive stars plays an essential role in determining stellar
structure at the time of their core collapse and thus their supernova (SN)
properties \citep[e.g.,][]{langer2012,yoon2015,vanbeveren2015,marchant2016}. In
particular, the lack of hydrogen-rich layers in progenitors of stripped-envelope
core-collapse SNe (i.e., Type~IIb/Ib/Ic SNe), is often suggested to be caused by
binary interaction
\citep[e.g.,][]{wheeler1985,ensman1988,podsi1992,nomoto1994,shigeyama1994,woosley1994,bersten2012,bersten2014,fremling2014,ergon2015,eldridge2015,lyman2014}.
The small typical ejecta mass estimated from light curves (LCs) of
stripped-envelope SNe ($\simeq 1-5~\Msun$, e.g.,
\citealt{sauer2006,drout2011,taddia2015,lyman2014}), and nucleosynthetic
signatures estimated from their spectral modeling
\citep[e.g.,][]{jerkstrand2015}, support progenitors with relatively small
zero-age main-sequence (ZAMS) masses. The less massive ZAMS mass stars need to
remove their hydrogen-rich envelopes with mass loss caused by binary
interactions because of their inefficient radiation-driven wind
\citep[e.g.,][]{podsi1992,nomoto1995,podsi2004,izzard2004,yoon2010,eldridge2011,benvenuto2013,lyman2014,eldridge2016}.
It is also known that mass loss caused by binary interaction is essential to
explain the observational ratio of stripped-envelope SNe to hydrogen-rich SNe
\citep[e.g.,][]{eldridge2008,eldridge2013,smith2011}.

Some Type~Ib/Ic SNe are known to have a much faster LC evolution than others.
While typical Type~Ib/Ic SNe reach their peak luminosity in $\sim 20$~days
\citep[e.g.,][]{drout2011,prentice2016}, rapidly-evolving SN LCs rise in less
than $\sim 10$~days and decline quickly on a similar timescale
\citep[e.g.,][]{poznanski2010,perets2010,kawabata2010,kasliwal2010,ofek2010,kasliwal2012,drout2013,drout2014,inserra2015}.
The simplest way to interpret the rapid LC evolution of some Type~Ib/Ic SNe is that
their ejecta mass is much smaller than in the more slowly evolving SNe (see
also, e.g., \citealt{kleiser2014,drout2014,tanaka2016}). LC evolution becomes
faster with smaller ejecta mass because of the smaller diffusion timescale. This
is roughly proportional to $(\Mej^3/\Eej)^{1/4}$, where $\Mej$ is the ejecta
mass and \Eej\ is the kinetic energy \citep[e.g.,][]{arnett1982}. The ejecta
mass of rapidly-evolving SNe is typically estimated to be $\sim 0.1~\Msun$,
which is an order of magnitude smaller than that of typical Type~Ib/Ic SNe
\citep[e.g.,][]{poznanski2010,kasliwal2012,drout2013}.

Type~Ib/Ic SNe with rapidly-evolving LCs show diversity in their peak luminosities and spectra. SN~2002bj is among the first observed rapidly-evolving Type~Ib/Ic SNe and among the brightest with its peak magnitude at around $-18$~mag \citep{poznanski2010}. Many rapidly-evolving SNe have their peak magnitudes at around $-16$~mag with different spectral properties. One example of rapidly-evolving SNe in this luminosity range is so-called ``Ca-rich gap transients.'' They are optical transients whose peak luminosity lie between that of classical novae and SNe and they show strong Ca lines especially in the nebular phase \citep[e.g.,][]{kasliwal2012}. Although they are called "Ca-rich" transients, they may not necessarily be Ca-rich. What is indicated from their spectra is that they have a larger fraction of Ca to O than typical Type~Ib/Ic SNe. Several progenitor scenarios have been suggested for these events \citep[e.g.,][]{perets2010,kawabata2010}, but their nature is not yet clear. There also exist rapidly-evolving SNe with similar luminosities to the Ca-rich gap transients but with very different spectral features such as SN~2005ek \citep{drout2013} and SN~2010X \citep{kasliwal2010}. Another kind of rapidly-evolving SNe are SN~2002cx-like SNe, which are also known as Type~Iax SNe; they are characterized by their strong Si and S features with relatively low photospheric velocities and by the wide peak luminosity range covering from $\sim -14$~mag to $\sim -19$~mag \citep[e.g.,][]{foley2013}.

A number of possibilities have been suggested to obtain a small ejecta
mass to explain the rapidly-evolving SNe, including some not related to the core collapse of massive stars  
\citep[e.g.,][]{moriya2016,dessart2015b,kashiyama2015,kleiser2014,shen2010,moriya2010,kitaura2006}.
In particuar, \citet{tauris2013,tauris2015} proposed an ``ultra-stripped''
core-collapse SN scenario to explain small ejecta masses. They showed that tight
helium star--neutron star (NS) binary systems, presumably created in the
common-envelope phase from high-mass X-ray binaries \citep{tauris2006}, can lead
to the extreme stripping of the helium envelope and result in SNe with ejecta
masses of the order of 0.1~\Msun\ or less. The SN ejecta mass from these systems
is even less than those typically obtained in SN progenitors from the first
exploding stars ($\sim 1~\Msun$) during binary evolution
(e.g., \citealt{yoon2010,lyman2014}).

Several studies have investigated the observational properties of
stripped-envelope SNe with ejecta larger than 1~\Msun\ coming from progenitors
obtained from binary stellar evolution \citep[e.g.,][]{dessart2015}. However,
few LC and spectral studies have been carried out for ultra-stripped SNe. A
previous study of ultra-stripped SNe \citep{tauris2013} only provided LC models.
Because there are several proposed ways to make SNe with rapidly-evolving LCs,
LC information is not sufficient to identify ultra-stripped SNe observationally.
In addition, the \Ni\ mass was treated as a free parameter in the previous
study. In this paper, we investigate not only LCs but also spectral properties
of ultra-stripped SNe by performing explosive nucleosynthesis calculations which
provide an appropriate estimate for the \Ni\ mass synthesized during the
explosion, so that we can have a better understanding of their observational
signatures and identify ultra-stripped SNe observationally.

Ultra-stripped SNe are closely connected to the formation of double NS systems.
It was argued by \citet{tauris2013,tauris2015} that {\it all} double NS systems
formed in the Galactic disk (i.e., outside dense environments such as globular
clusters) which are tight enough to merge within a Hubble time {\it must} have
been produced from an ultra-stripped SN. Therefore, ultra-stripped SNe are
related to sources of gravitational waves detected by LIGO
\citep[e.g.,][]{abadie2010}. In addition, double NS systems which merge are
suggested to cause short gamma-ray bursts
\citep{blinnikov1984,paczynski1986,narayan1992} and $r$-process element
nucleosynthesis \citep[e.g.,][]{rosswog1999,argast2004,hirai2015}, possibly followed by
a so-called ``kilonova'' \citep[e.g.,][]{metzger2010,barnes2013,tanaka2013}.
However, to produce a double NS system a massive binary must survive two SN
explosions. 
A binary system is likely to be disrupted by the first SN explosion
because of a combination of sudden mass loss and a kick imparted to the newborn
NS \citep{brandt1995}. Ultra-stripped SNe, on the other hand, eject
very little mass and are unlikely to have large NS kicks in general because of
their rapid explosions \citep{podsi2004b,suwa2015,tauris2015}. 
Thus, ultra-stripped SNe avoid these two major
obstacles in forming double NS systems and are therefore likely to produce
systems which lead to merger events if the post-SN orbital period is short
enough.
Future constraints on the observed rate of ultra-stripped SNe can be
directly compared to the double NS
merger rate determined from gravitational wave observations \citep{tauris2015}.
This can be used to verify their evolutionary connection.
The merger rate of double NS systems will be known within a few years when
LIGO/VIRGO reach full design sensitivity. We refer to \citet{abadie2010,berry2015,abbott2016b}
for detailed reviews on the expected merger rates.

This paper is organized as follows. Section~\ref{sec:method} presents the
numerical methods we adopt in our study. Our synthetic LC and spectral models
are presented in Section~\ref{sec:results}. We compare our results with
observations in Section~\ref{sec:compobs}. We discuss our results in
Section~\ref{sec:discussion} and conclude this paper in
Section~\ref{sec:conclusions}.

\begin{figure}
 \begin{center}
  \includegraphics[width=\columnwidth]{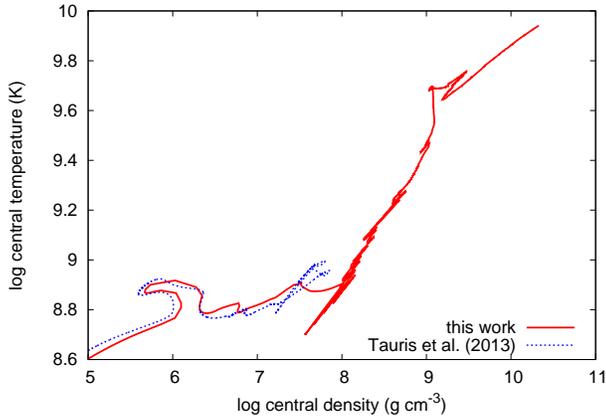} 
 \end{center}
\caption{
Evolution of the central density and temperature of our ultra-stripped SN progenitor obtained with \texttt{MESA}. We also present the central density and temperature evolution obtained by \citet{tauris2013} with \texttt{BEC}.
}\label{fig:center}
\end{figure}

\begin{figure}
 \begin{center}
  \includegraphics[width=\columnwidth]{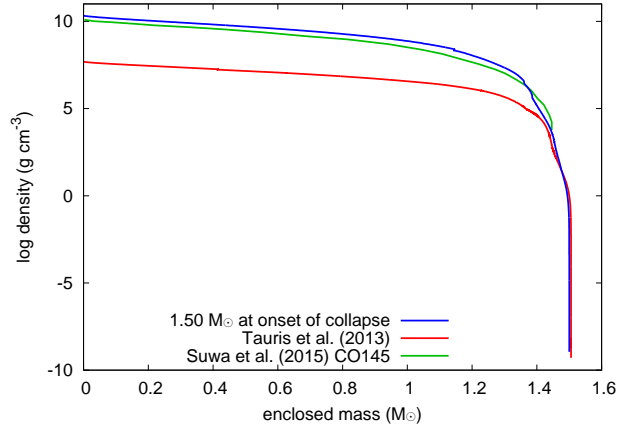} 
 \end{center}
\caption{
Progenitor density structure at core collapse. We also show the density 
structure from the model of \citet{tauris2013} which is shortly after 
off-center oxygen ignition, some $\sim\!10\;{\rm years}$ prior to core collapse. We show the density structure at the collapse from \citet{suwa2015} from the progenitor having the same carbon+oxygen core mass of 1.45~\Msun\ as our model.
}\label{fig:progenitordensity}
\end{figure}

\section{Method}\label{sec:method}
\subsection{Progenitor}

We investigate the observational properties of explosions originating from the
ultra-stripped SN progenitor presented in \citet{tauris2013} (see also
\citealt{tauris2015}). This is a typical model of ultra-stripped SNe. The total
progenitor mass at explosion is 1.50~\Msun\ with a carbon+oxygen core of
1.45~\Msun. The helium ZAMS mass of the progenitor was 2.9~\Msun.
\citet{tauris2013} evolved
the progenitor until shortly after the onset of off-center oxygen burning using
the \texttt{BEC} binary stellar evolution code (\citealt{yoon2010} and
references therein). Although the remaining time to collapse is estimated to be
about 10 years and the final progenitor mass does not change until the collapse,
the density in the inner layers of the progenitor increases significantly in the
remaining time. As the final density structure is critical for the explosive
nucleosynthesis results, we need to follow the progenitor evolution until core
collapse. The \texttt{BEC} code is not suitable for this.

In order to obtain a SN progenitor evolved until core collapse, we therefore used
the public stellar evolution code \texttt{MESA}
\citep{paxton2011,paxton2013,paxton2015}. We used the same physical parameters
as in the \citet{tauris2013} \texttt{BEC} calculations in \texttt{MESA}. In
particular, we used a mixing length parameter of 2 with a semiconvection efficiency parameter of 1.
As the core mass is rather small, weak reactions play an important role in the
evolution of the core approaching collapse
\citep[e.g.,][]{takahashi2013,jones2013,schwab2015}. Therefore we used the large
nuclear network \texttt{``mesa151.net''} provided in \texttt{MESA}, which
includes 151 nuclei up to $^{65}$Ni with important weak reactions such as
electron capture by $^{33}$S and $^{35}$Cl. We stopped the calculation when the
stellar core started to infall with a speed of 1,000~$\mathrm{km~s^{-1}}$.

First, we evolved a star starting from the helium ZAMS as is done by
\citet{tauris2013}, but as a single star. Because the mass lost by the binary
interactions like Roche-lobe overflow
is known from the binary calculation of \citet{tauris2013}, we
imposed their mass loss in our single-star evolution calculation. Thus, our
single-star evolution calculation takes mass loss caused by
binary stellar evolution into account in a simplified way.
\cite{tauris2013} obtained a 1.50~\Msun\ helium star with a
carbon+oxygen core of 1.45~\Msun\ from a helium ZAMS star of 2.90~\Msun\ which
suffers from Roche-lobe overflow to a NS with initial orbital period of
0.1~days. We slightly increased the initial helium mass to 2.949~\Msun\ in order
to obtain the same carbon+oxygen core mass after the core helium burning in our
\texttt{MESA} calculation. 
The evolution of the central density and temperature is presented in Fig.~\ref{fig:center}.
We find that the internal evolution of our progenitor is essentially the same as
that in \citet{tauris2013} up to the point they succeeded to follow.
The star forms an Fe core and collapses
as was presumed by \citet{tauris2013,tauris2015}.
It is suggested that low-mass core-collapse SN progenitors experience violent silicon
flashes shortly before the core collapse \citep{woosley2015}, but we do not find such
a flash in our model. If such a flash occurs, it may result in a creation of a dense
helium-rich circumstellar medium around the progenitor and the SN may be observed
as Type Ibn \citep[e.g.,][]{moriya2016maeda}.

The final density structure of the progenitor at collapse is presented in
Fig.~\ref{fig:progenitordensity}. The abundance of the Fe-group elements sharply
increases at an enclosed mass of 1.35~\Msun, while the electron fraction sharply
decreases there. Thus, we estimate that the final iron-core mass at the time of
collapse is 1.35~\Msun. This iron-core mass is comparable to that obtained
recently by \citet{suwa2015} (1.33~\Msun) from a progenitor model with a similar
carbon+oxygen core mass as ours (1.45~\Msun). We compare our progenitor with
that of \citet{suwa2015} in Fig.~\ref{fig:progenitordensity}. The internal
structure of the collapsing models is similar. We also show for comparison the
density structure obtained with the \texttt{BEC} code at a time $\sim\!10\;{\rm
years}$ prior to core collapse.

\subsection{Nucleosynthesis}

The collapsing progenitor described in the previous section was then used to
follow the explosive nucleosynthesis. Numerical calculations of explosive
nucleosynthesis were performed with the same numerical code as in previous
studies of explosive nucleosynthesis in core-collapse SNe
\citep[e.g.,][]{nakamura2001,tominaga2007}. It is a one-dimensional explicit
Lagrangian hydrodynamics code in which a piece-wise parabolic method is adopted
\citep{colella1984}. The $\alpha$-network is coupled with hydrodynamics and
detailed nucleosynthesis calculations are performed as post-processing
modelling. We used a reaction network including 280 isotopes up to $^{79}$Br
(Table~1 in \citealt{umeda2005}).

\subsection{Explosive hydrodynamics and light curves}

Synthetic LCs were numerically obtained using the one-dimensional multi-group
radiation hydrodynamics code \texttt{STELLA}
\citep{blinnikov1993,blinnikov1998,blinnikov2004,blinnikov2006,baklanov2005,sorokina2015}.
\texttt{STELLA} has been used to model SN LCs of various kinds, including
ultra-stripped SNe \citep{tauris2013}. We take the progenitor structure above a
mass cut and inject thermal energy at the bottom of the structure to initiate
the explosion. The amount of thermal energy injected is $\Eej+E_\mathrm{bind}$,
where $E_\mathrm{bind}$ is the total binding energy of the progenitor. We
used the chemical composition obtained from explosive nucleosynthesis.

We show LC and spectral models obtained using two different mass cuts, at 
1.30~\Msun\ and 1.35~\Msun, respectively. A mass-cut of 1.35~\Msun\ corresponds
to the final iron-core mass of the progenitor model. \citet{suwa2015} obtained a
final NS baryonic mass of 1.35~\Msun\ from an explosion using the same
carbon+oxygen core mass. \cite{tauris2013} used a mass cut of 1.30~\Msun. Both
the small Fe core mass of our progenitor model and simulations by
\citet{suwa2015} suggest that the mass cut is likely to be small. 

\begin{table*}
\centering
\caption{Explosive nucleosynthesis results of the $1.50~\Msun$ model for 
different ejecta masses and explosion energies. Mass fractions at the end of 
the nucleosynthesis calculations are presented.
The value in each column, $a$ $(x)$, means that the corresponding mass fraction 
is $a\times 10^x$. 
The abundance estimated for SN~2005ek based on its peak spectrum 
\citep{drout2013} is also shown.
$T_\mathrm{max}$ is the maximum temperature reached during the explosive 
nucleosynthesis calculations.}
\label{table:abn}
\begin{tabular}{cccccccc}
\hline
element  & \multicolumn{3}{c}{$\Mej=0.20~\Msun$} & \multicolumn{3}{c}{$\Mej=0.15~\Msun$} & SN 2005ek \\
  & 0.50 B & 0.25 B & 0.10 B & 0.50 B & 0.25 B & 0.10 B &  \\
\hline
He  & 2.4 $(-1)$ & 2.3 $(-1)$ & 2.2 $(-1)$ & 3.0 $(-1)$ & 2.9 $(-1)$ & 2.8 $(-1)$ & - \\
C   & 4.3 $(-2)$ & 4.5 $(-2)$ & 4.7 $(-2)$ & 5.8 $(-2)$ & 6.0 $(-2)$ & 6.3 $(-2)$ & 2.0 $(-2)$\\
O   & 1.7 $(-1)$ & 1.8 $(-1)$ & 1.9 $(-1)$ & 2.3 $(-1)$ & 2.4 $(-1)$ & 2.5 $(-1)$ & 8.6 $(-1)$\\
Ne  & 5.2 $(-2)$ & 6.9 $(-2)$ & 7.9 $(-2)$ & 7.0 $(-2)$ & 9.2 $(-2)$ & 1.1 $(-1)$ & - \\
Mg  & 2.3 $(-2)$ & 2.5 $(-2)$ & 2.6 $(-2)$ & 3.0 $(-2)$ & 3.4 $(-2)$ & 3.5 $(-2)$ & 8.2 $(-2)$\\
Si  & 5.1 $(-2)$ & 5.3 $(-2)$ & 6.0 $(-2)$ & 6.7 $(-2)$ & 7.0 $(-2)$ & 8.0 $(-2)$ & 2.5 $(-2)$\\
S   & 2.4 $(-2)$ & 2.5 $(-2)$ & 2.3 $(-2)$ & 3.2 $(-2)$ & 3.4 $(-2)$ & 4.0 $(-2)$ & 8.0 $(-3)$\\
Ca  & 3.9 $(-3)$ & 4.4 $(-3)$ & 5.1 $(-3)$ & 5.2 $(-3)$ & 5.8 $(-3)$ & 6.7 $(-3)$ & 1.2 $(-3)$\\
Sc  & 1.0 $(-6)$ & 1.0 $(-6)$ & 3.4 $(-7)$ & 4.3 $(-7)$ & 1.1 $(-6)$ & 3.8 $(-7)$ & -\\
Ti  & 1.3 $(-4)$ & 7.2 $(-5)$ & 2.8 $(-5)$ & 1.3 $(-4)$ & 7.8 $(-5)$ & 3.1 $(-5)$ & 3.3 $(-4)$\\
V   & 3.2 $(-6)$ & 5.8 $(-6)$ & 2.4 $(-6)$ & 2.4 $(-6)$ & 6.6 $(-6)$ & 2.8 $(-6)$ & -\\
Cr  & 2.7 $(-4)$ & 2.2 $(-4)$ & 1.9 $(-4)$ & 3.0 $(-4)$ & 2.6 $(-4)$ & 2.4 $(-4)$ & 3.3 $(-4)$\\
Mn  & 1.9 $(-5)$ & 2.6 $(-5)$ & 2.8 $(-5)$ & 2.4 $(-5)$ & 3.4 $(-5)$ & 3.8 $(-5)$ & - \\
Fe  & 3.3 $(-3)$ & 4.3 $(-3)$ & 6.0 $(-3)$ & 4.3 $(-3)$ & 5.7 $(-3)$ & 8.0 $(-3)$ & 1.5 $(-3)$ \\
Co  & 2.8 $(-4)$ & 4.0 $(-4)$ & 5.8 $(-4)$ & 3.7 $(-4)$ & 5.3 $(-4)$ & 7.7 $(-4)$ & 1.2 $(-3)$ \\
\hline
\Ni\ mass  & 0.034~\Msun & 0.030~\Msun & 0.026~\Msun & 0.031~\Msun & 0.027~\Msun & 0.021~\Msun & 0.03~\Msun\\
$T_\mathrm{max}/10^9$ K & 14 & 12 & 11 & 7.5 & 6.6 & 6.0 & -\\
\hline
\end{tabular}
\end{table*}

\subsection{Spectra}

The spectral properties of ultra-stripped SNe have been investigated using the
Monte Carlo spectral synthesis code developed by \citet{mazzali1993}. We refer
to \citet{mazzali1993,lucy1999,mazzali2000b,tanaka2011} for details. This code
has been used for many SN spectral synthesis studies
\citep[e.g.,][]{mazzali1992,mazzali1993b,mazzali2000,tanaka2011}.

The code is applicable in early phases of SNe when a photosphere exists in the
ejecta from where photons are assumed to be emitted with a blackbody spectrum.
The code requires a density structure, abundances, position of the photosphere,
and emerging SN luminosity to synthesize spectra. We used the average abundances
of the models obtained from the nucleosynthesis calculations
(Table~\ref{table:abn}) in our spectral modelling. We did not assume
stratification of chemical elements in the SN ejecta because the ejecta mass is
small and the ejecta are likely to be well mixed \citep[e.g.,][]{hachisu1991}. We took the
density structure and luminosity from \texttt{STELLA}. The spectral code assumes
homologous expansion of the SN ejecta, which is satisfied in every model we
present in this paper. A converging model is obtained by changing photospheric
velocity and temperature in the spectral synthesis code. 

Our models contain a large fraction of helium, and some rapidly-evolving SNe are
of Type~Ib. However, the code we mainly use does not include non-thermal
excitation of helium which is essential in modelling helium features in SN
spectra \citep[e.g.,][]{lucy1991,mazzalilucy98,hachinger2012}. We investigate
the effect of non-thermal helium excitation using a code developed by
\citet{hachinger2012} and find that non-thermal excitation does not have a
strong effect on the spectra we present in this study
(Section~\ref{sec:hefeatures}).

\section{Results}\label{sec:results}
\subsection{Nucleosynthesis}

We calculated explosive nucleosynthesis for three different explosion energies:
0.10, 0.25, and 0.50~B $(1~\mathrm{B}\equiv 10^{51}~\mathrm{erg})$. We
investigated these small explosion energies based on the recent explosion
simulations of ultra-stripped SNe by \citet{suwa2015}. They found that
ultra-stripped SNe explode via the neutrino-driven mechanism and have
small explosion energies, of the order of 0.1~B.

\begin{figure}
 \begin{center}
  \includegraphics[width=\columnwidth]{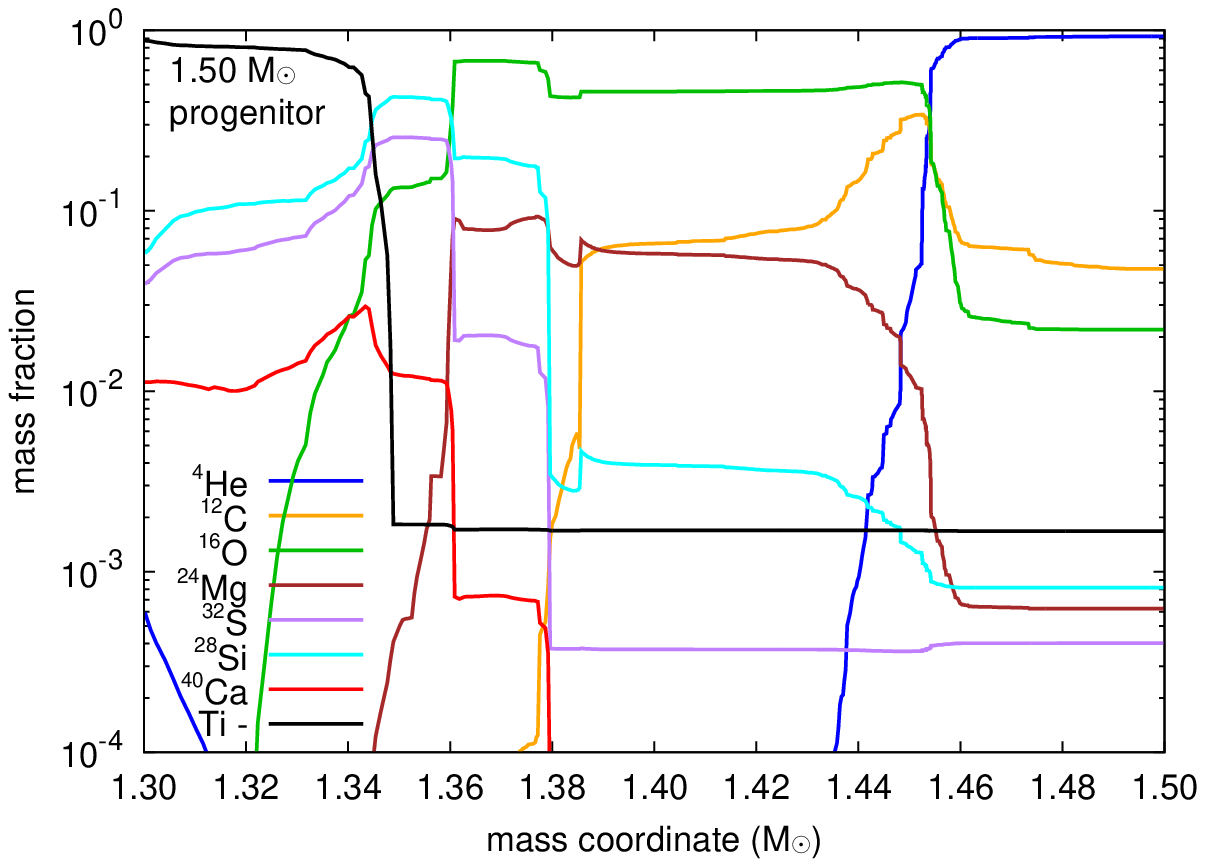}
  \includegraphics[width=\columnwidth]{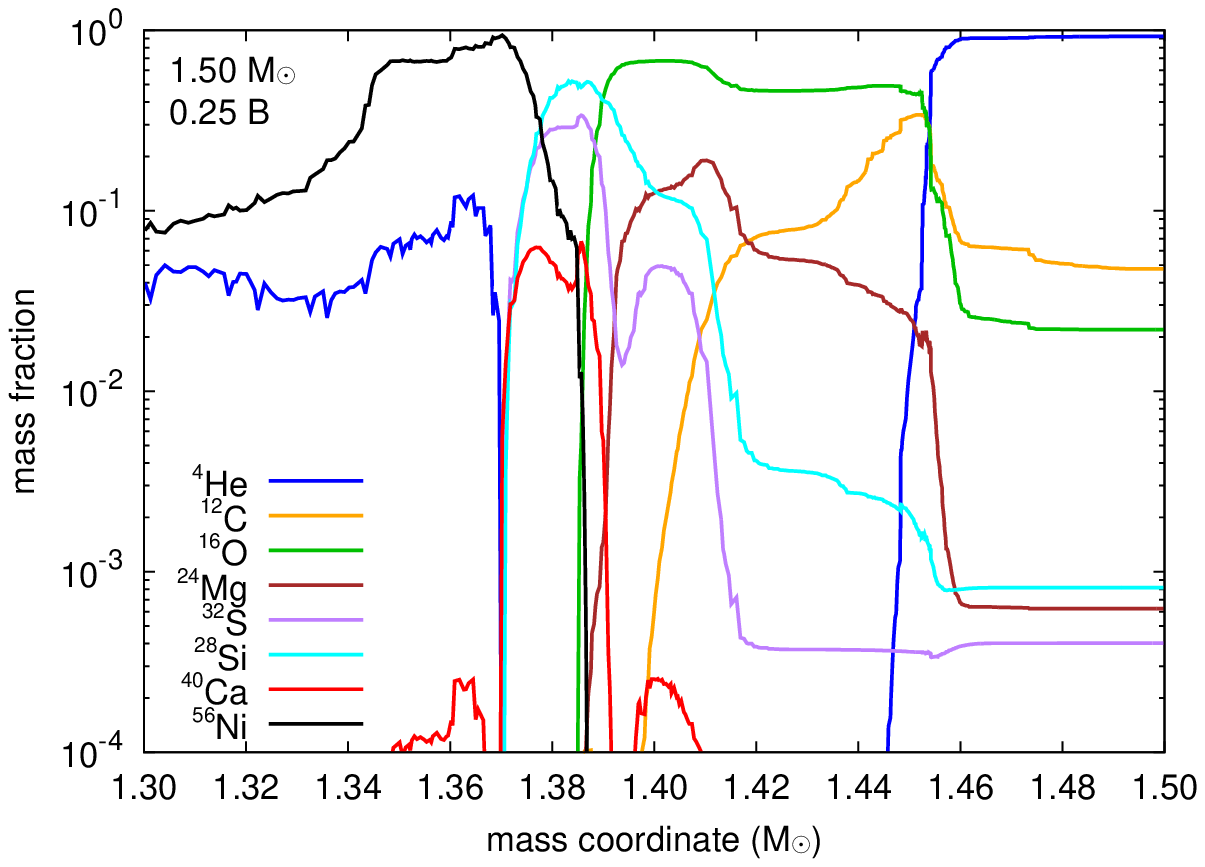} 
 \end{center}
\caption{
Chemical structure of the 1.50~\Msun\ progenitor at iron core collapse (top) and 
the result of our explosive nucleosynthesis calculation with the explosion 
energy 0.25~B (bottom). The total abundance above Ti is plotted as a single line in the top panel (``Ti -'').
}\label{fig:nuc}
\end{figure}

The results of our nucleosynthesis calculation for the case of
$\Eej=0.25~\mathrm{B}$ is shown in Fig.~\ref{fig:nuc}. Table~\ref{table:abn}
shows the final average abundances in the ejecta for all energies and mass cuts
we applied. One of the most important elements determining SN properties is the
mass of \Ni\ synthesised in the explosion.
We find that the \Ni\ masses range from 0.034 to 0.021~\Msun\ depending on explosion energy and mass cut (Table~\ref{table:abn}).

\subsection{Light curves}

\begin{figure}
 \begin{center}
  \includegraphics[width=\columnwidth]{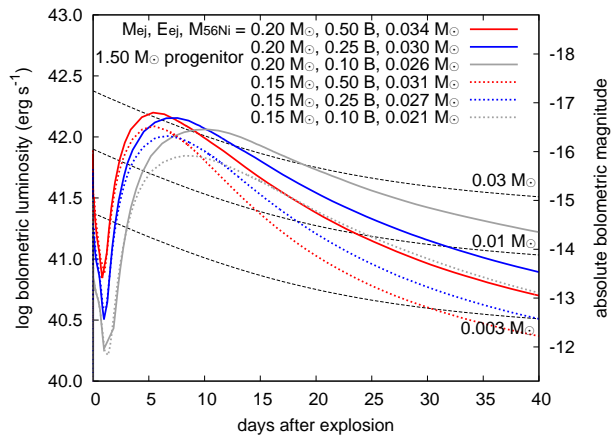} 
 \end{center}
\caption{
Bolometric LCs of ultra-stripped SNe.
The thin dashed black lines show the total available energy from the nuclear 
decay of the indicated \Ni\ masses.
}\label{fig:lcb}
\end{figure}

Figure~\ref{fig:lcb} shows the bolometric LCs of ultra-stripped SNe obtained in
this study. After shock breakout, the bolometric LCs decline in the first 1~day
owing to adiabatic cooling of the ejecta. Then, when the heating from \Ni\ decay
becomes dominant, the bolometric luminosity starts to increase again. The peak
luminosity is approximately proportional to the initial \Ni\ mass. In most
cases, the peak luminosity does not exactly match that expected from a simple
estimate based on \citet{arnett1982}: it is larger by up to 50 per cent. The rise time
and peak luminosity are consistent with those estimated in \citet{tauris2015} by
using an analytic approach.

\begin{figure}
 \begin{center}
  \includegraphics[width=\columnwidth]{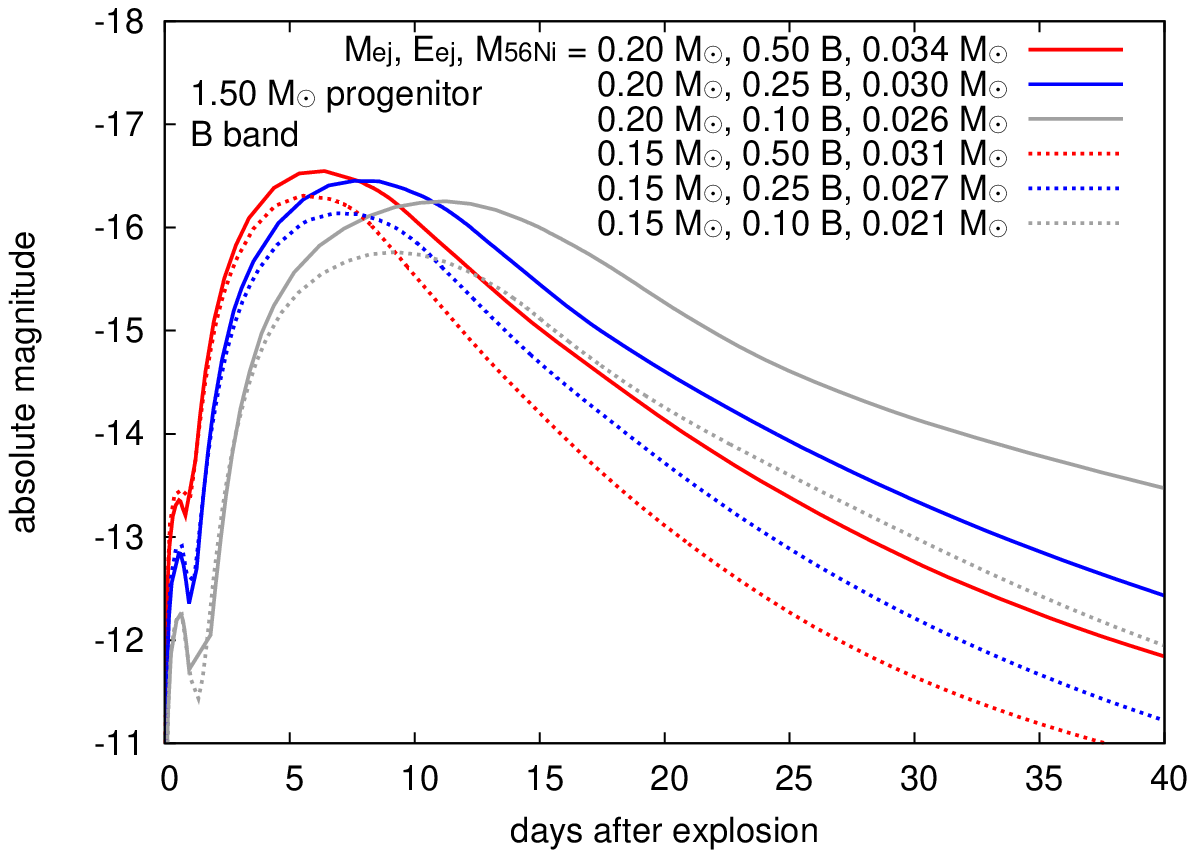} \\
  \includegraphics[width=\columnwidth]{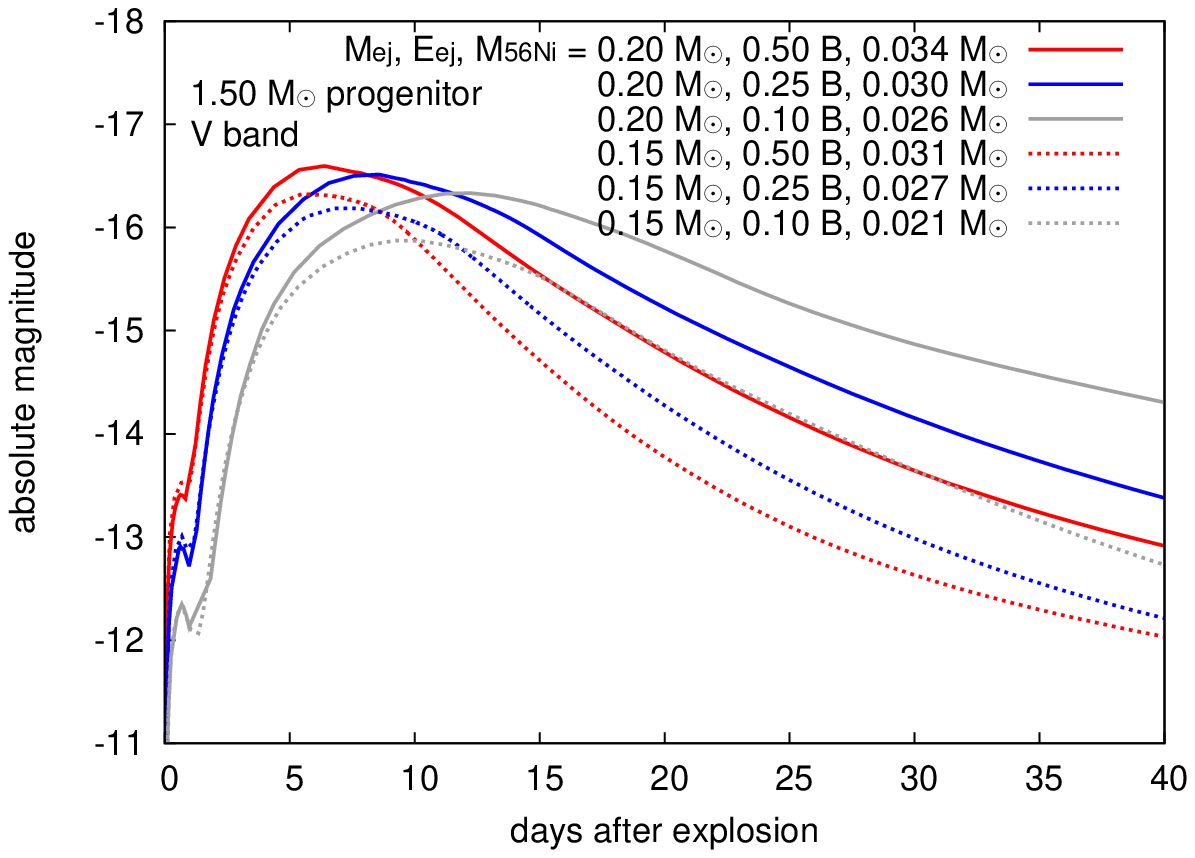} \\
  \includegraphics[width=\columnwidth]{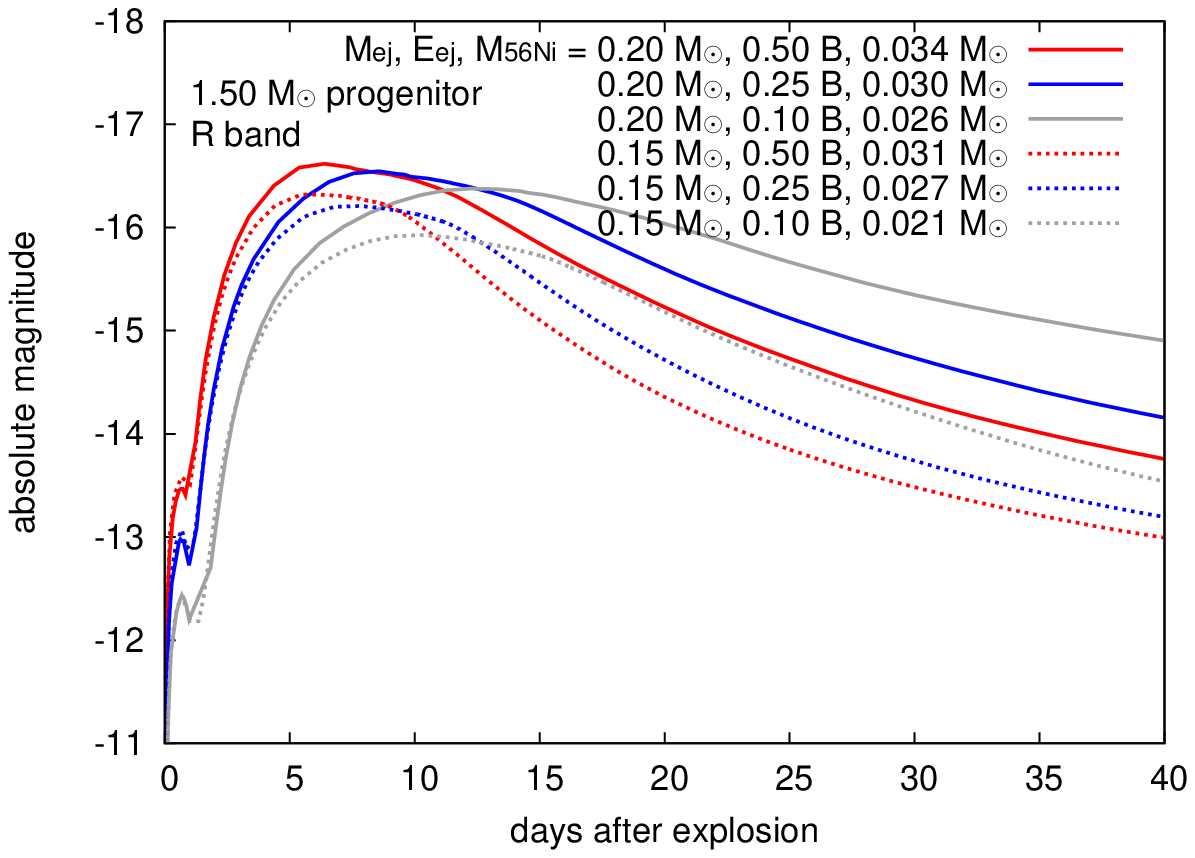} 
 \end{center}
\caption{
Multi-color LCs for the models shown in Fig.~\ref{fig:lcb}.
}\label{fig:lcc}
\end{figure}

Figure~\ref{fig:lcc} shows multi-color LCs of the same models with several
Bessel filters \citep{bessel1990}. Optical LCs presented in Fig.~\ref{fig:lcc}
show some differences from the bolometric LCs. In particular, optical LCs show a first
LC peak at times when bolometric LCs monotonically decline. This is a
consequence of the cooling of the ejecta, which shifts the spectral peak to
longer wavelengths as time goes on. Therefore, LCs in redder bands peak later.

\begin{figure}
 \begin{center}
  \includegraphics[width=\columnwidth]{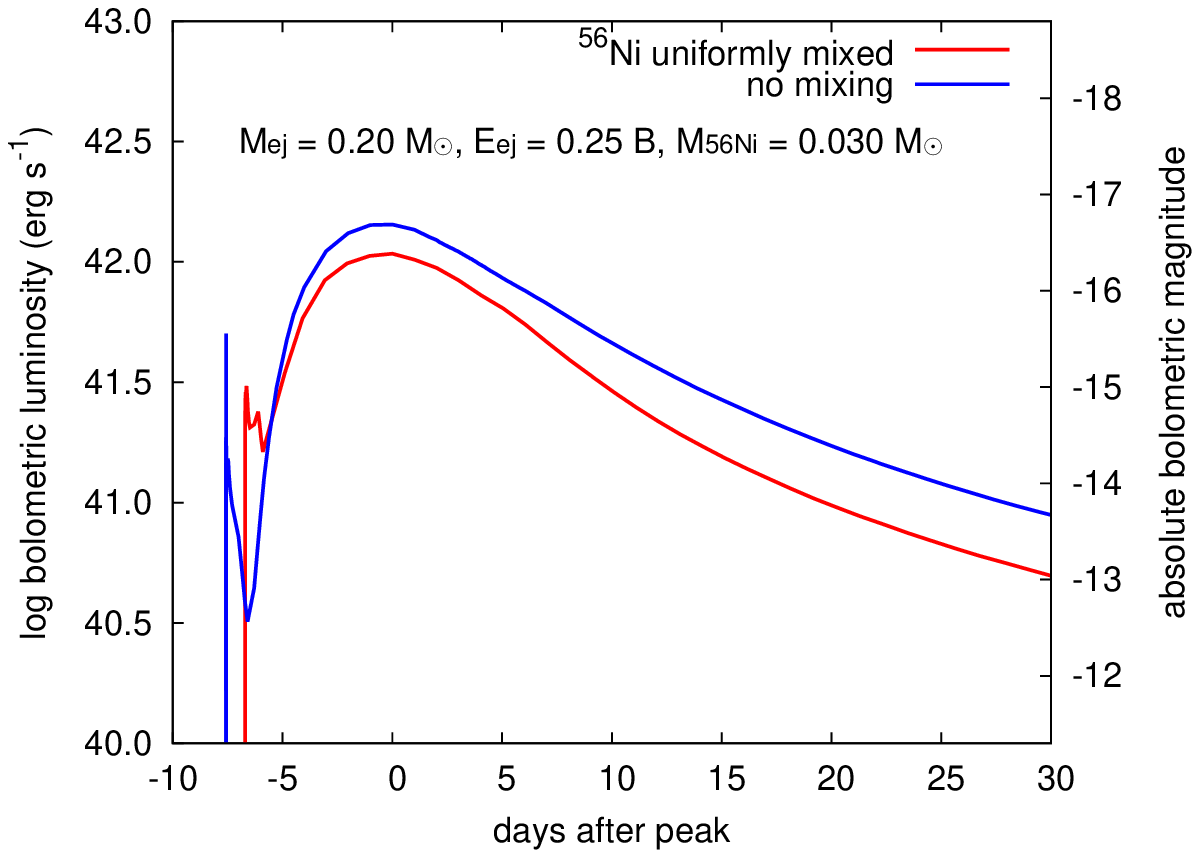} 
 \end{center}
\caption{
Bolometric LCs with and without \Ni\ mixing. Bolometric LCs from the models 
with $\Mej=0.20~\Msun$ and $\Eej=0.25~\mathrm{B}$ are shown.
}\label{fig:lcmix}
\end{figure}

In the models presented so far, we have used the chemical structure from the
explosive nucleosynthesis modeling (cf. Fig.~\ref{fig:nuc}) and did not take
into account the effect of mixing. To demonstrate the effect of mixing on the
LCs, we show a LC in which \Ni\ is uniformly mixed in the entire ejecta
(Fig.~\ref{fig:lcmix}). Because of the presence of \Ni\ in the outer layers,
heating by \Ni\ in ejecta is more efficient early on and the rise time becomes
shorter in the mixed model. However, at late phases, the gamma-rays in the outer
layers are less trapped because of the smaller optical depth. Thus, the
luminosity of the mixed model is less than that of the non-mixed model by about
50\%. The decline rate after the LC peak is not strongly affected by mixing.

\subsection{Spectra}
\subsubsection{Peak spectra}
\begin{figure}
 \begin{center}
  \includegraphics[width=1\columnwidth]{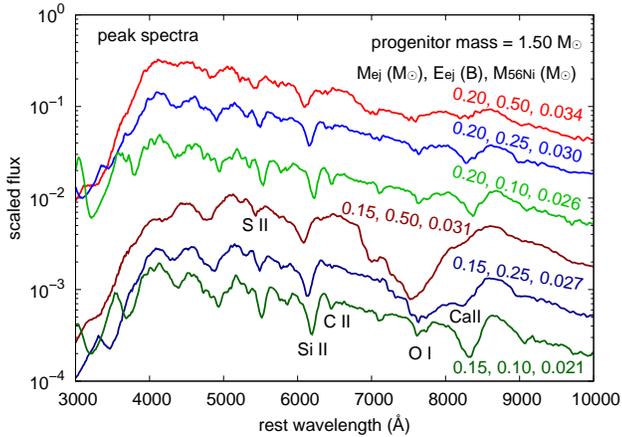}
 \end{center}
\caption{
Synthetic spectra at bolometric LC peak for various models.
The time of the bolometric LC peak is (from the top spectrum to the bottom) 
5.3~days, 7.5~days, 10.2~days, 5.4~days, 6.8~days, and 8.2~days.
}\label{fig:peaksp}
\end{figure}

We present synthetic spectra at maximum light in Fig.~\ref{fig:peaksp}. We focus
on this epoch's spectra because this is when ultra-stripped SNe are most likely to
be observed. We compare these synthetic spectra with observations in
Section~\ref{sec:spcomp}.

\begin{figure}
 \begin{center}
  \includegraphics[width=1\columnwidth]{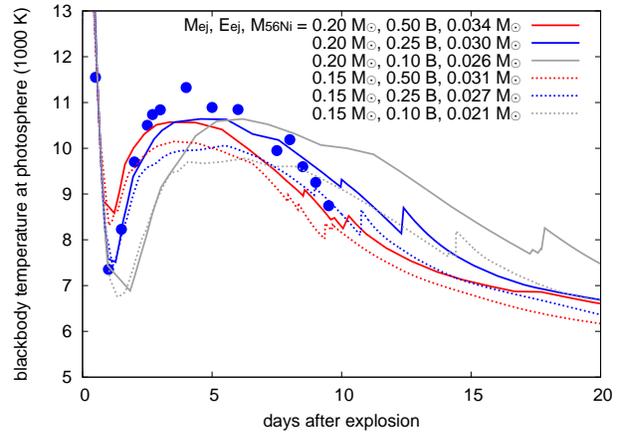} \\
  \includegraphics[width=1\columnwidth]{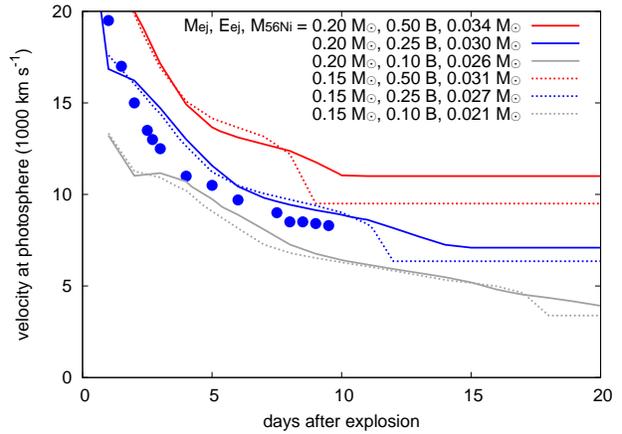}
 \end{center}
\caption{
Photospheric temperature and velocity of the models from our \texttt{STELLA} LC 
calculations (solid and dotted lines). The filled circles indicate the results 
of the spectral synthesis code using the density structure from \texttt{STELLA} 
for the $\Mej = 0.20~\Msun$, $\Eej=0.25~\mathrm{B}$ model. The synthesized 
spectra are presented in Fig.~\ref{fig:spseq}.
}\label{fig:photo}
\end{figure}

Spectral features primarily depend on elemental abundances, photospheric
temperature, and photospheric velocity. We used the mixed composition in
Table~\ref{table:abn} for our spectral modelling. Figure~\ref{fig:photo} shows
the photospheric velocities and temperatures obtained by fitting the blackbody
function to the spectral energy distributions (SEDs) obtained from
\texttt{STELLA}. The photosphere here is defined as the location where the
Rosseland mean optical depth is 2/3. The circles in Fig.~\ref{fig:photo}
represent the values in the converged spectral models of $\Mej=0.20~\Msun$ and
$\Eej=0.25~\mathrm{B}$ shown in Fig.~\ref{fig:spseq}. The photospheric
temperature and velocity from the LC code and the spectral code match within 10
per cent.

Line shifts and broadening in spectra depend on the photospheric velocity.
Because velocity is proportional to $(\Eej/\Mej)^{1/2}$, $\Eej/\Mej$ is a good
indicator of these properties. The models with $\Eej=0.25~\mathrm{B}$ have
$\Eej/\Mej\simeq 1~\mathrm{B}/\Msun$, which is similar to typical SNe~Ia.

There are several notable features in our synthetic peak spectra. First of all,
there are relatively strong Si~\textsc{ii} features, particularly Si~\textsc{ii}
$\lambda 6355$. In some models, especially in the spectra with relatively small
explosion energy, we can also see the C~\textsc{ii} $\lambda 6582$ feature next
to Si~\textsc{ii} $\lambda 6355$. We can find some S~\textsc{ii} features
between 5000~\AA\ and 6000~\AA. O~\textsc{i} $\lambda 7774$ and Ca~\textsc{ii}
IR triplet around 8000~\AA\ are also seen. The strong Ca feature in the
$\Mej=0.15~\Msun$ and $\Eej=0.50~\mathrm{B}$ model is due to Ca ionization as is
discussed in the next section.

\subsubsection{Temporal evolution}

\begin{figure}
 \begin{center}
  \includegraphics[width=\columnwidth]{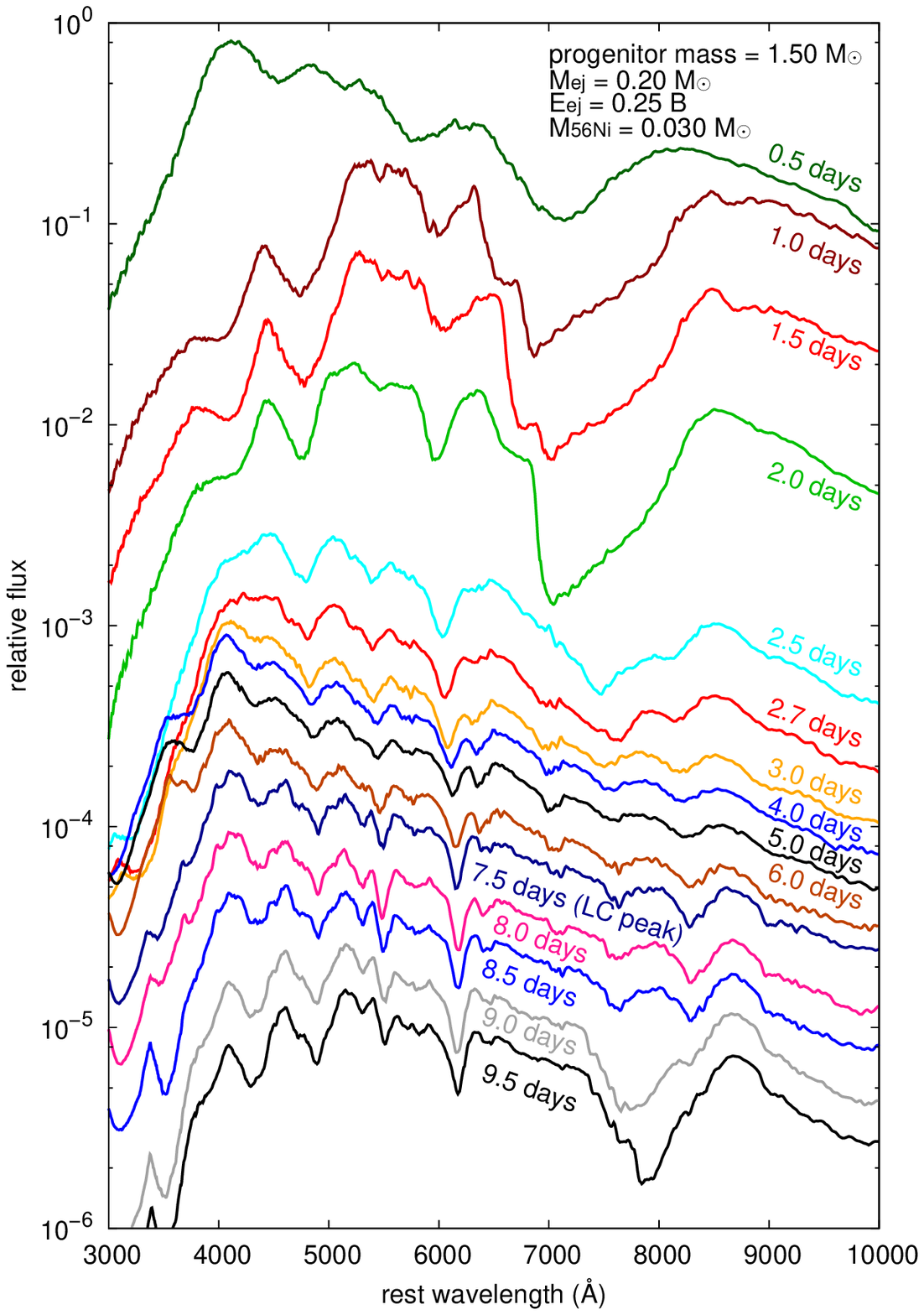} 
 \end{center}
\caption{
Temporal evolution of spectra for the model with $\Mej=0.20~\Msun$ and
$\Eej =0.25~\mathrm{B}$.
}\label{fig:spseq}
\end{figure}

We show the temporal evolution of our synthetic spectra with $\Mej=0.20~\Msun$
and $\Eej=0.25~\mathrm{B}$ in Fig.~\ref{fig:spseq}. We first present the
spectrum at 0.5~days after the explosion, which is still in the cooling phase
after shock breakout (Fig.~\ref{fig:lcb}). This epoch roughly corresponds to the
time when the optical LCs show the first peak (Fig.~\ref{fig:lcc}). The spectra
during these very early epochs are similar to those of broad-line Type Ic SNe
(SNe Ic-BL) because the photosphere is located in the high-velocity outer layers
at that time (cf. Fig.~\ref{fig:photo}).

Up to about 2.5~days after the explosion, a strong Ca absorption is observed
between 7000~\AA\ and 8000~\AA. These broad features disappear by the time of LC
peak, when the photospheric temperature becomes hot enough to change the Ca
ionization level. The broad feature starts to appear again about 9~days after
the explosion as the photosphere cools. There is no significant evolution in the
spectra around LC peak, and the spectra quickly evolve to become nebular soon
thereafter. A further study of the late-time spectra is beyond the scope of this
paper. 

\subsubsection{Helium features}\label{sec:hefeatures}

Our progenitor contains about 0.03~\Msun\ of helium. This is close to the
maximum helium mass that can be hidden without showing significant spectral
features in relatively low-mass SNe~Ib/Ic \citep[about
0.1~\Msun\ of He][]{hachinger2012}. It is important to judge whether ultra-stripped SNe
from our progenitor are expected to be observed as Type~Ib or Type~Ic. However,
the spectral synthesis code we have used so far does not take the non-thermal
excitation required for helium excitation into account. Here, we show a spectral
model where the effect of the non-thermal excitation is taken into account. The
spectrum  was computed with the spectral synthesis code described in
\citet{hachinger2012}.

\begin{figure}
 \begin{center} 
  \includegraphics[width=1\columnwidth]{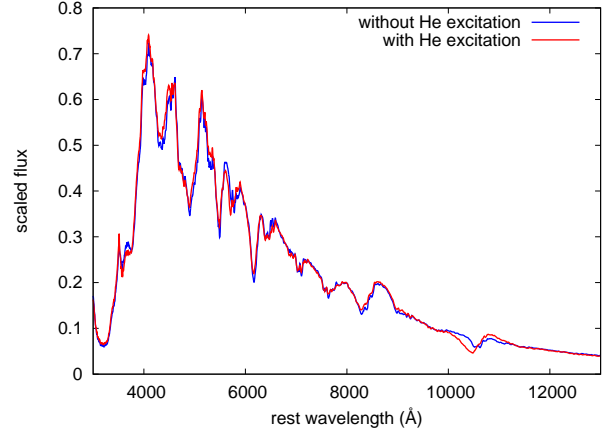}
 \end{center}
\caption{
Synthetic spectra with with and without non-thermal helium excitation. The 
original model is the maximum light spectrum of the $\Mej=0.20~\Msun$, 
$\Eej=0.25~\mathrm{B}$ model with $\Mni=0.025~\Msun$, 
which is also shown in Fig.~\ref{fig:peaksp}.
}\label{fig:diffheexcart}
\end{figure}

Figure~\ref{fig:diffheexcart} shows the spectrum including non-thermal
excitation. In the optical range the spectrum is not strongly affected by
non-thermal effects. Helium lines remain invisible except for the intrinsically
strong $\lambda 10830$ multiplet. Therefore, non-thermal excitation does not
seem to cause significant changes to the spectra of ultra-stripped SNe.

The fact that He\,\textsc{i} remains practically invisible in our spectra, while
0.1\,\Msun\ of helium could be detected in the SN Ib/Ic models of
\citet{hachinger2012}, is not only due to the somewhat smaller amount of \Ni\ in
our models. It can also be traced back to the fact that spectral temperatures
are relatively high in our models, such that the non-thermal excitation effects
rather result in a high ionization fraction (i.e. dominance of He\,\textsc{ii})
than in a large occupation number within He\,\textsc{i} excited states possibly
generating lines. This is a situation somewhat similar to that of SLSNe
\citep{mazzali2016}.

\begin{figure}
 \begin{center}
  \includegraphics[width=\columnwidth]{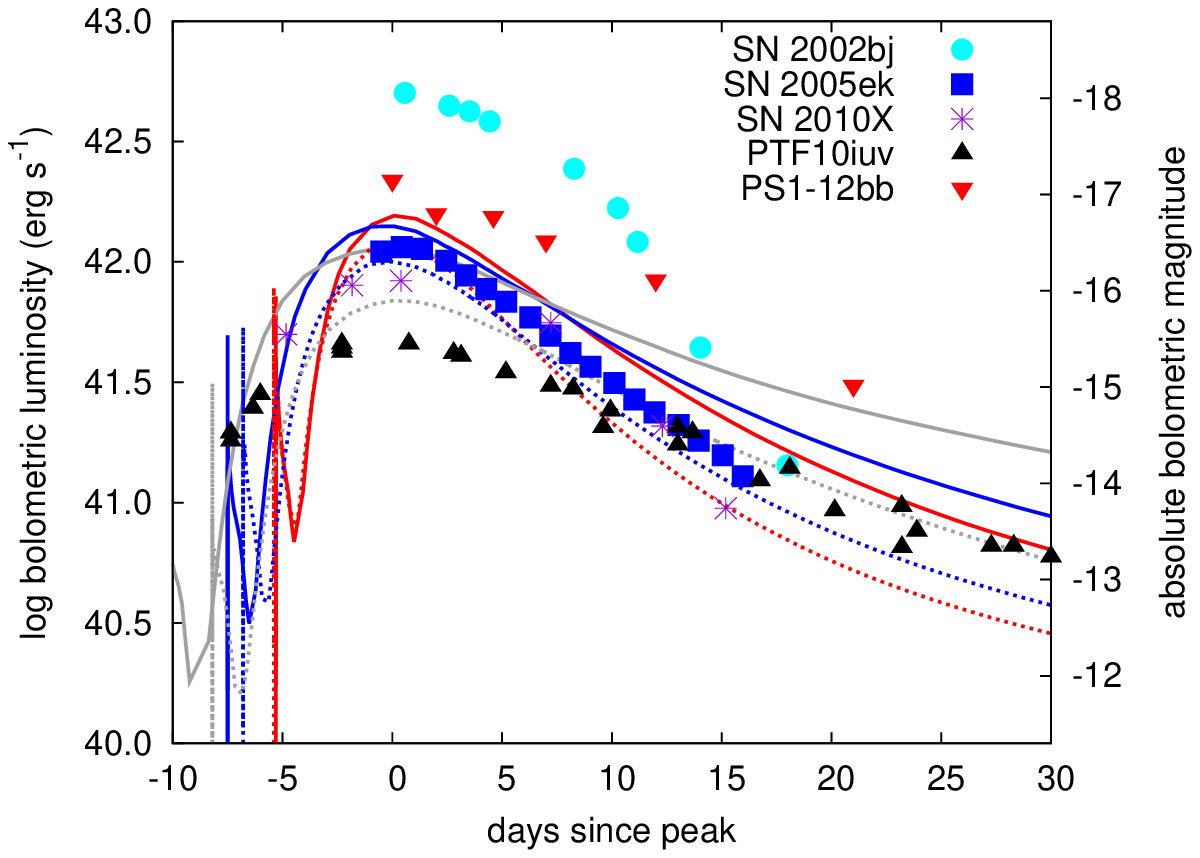} 
  \includegraphics[width=\columnwidth]{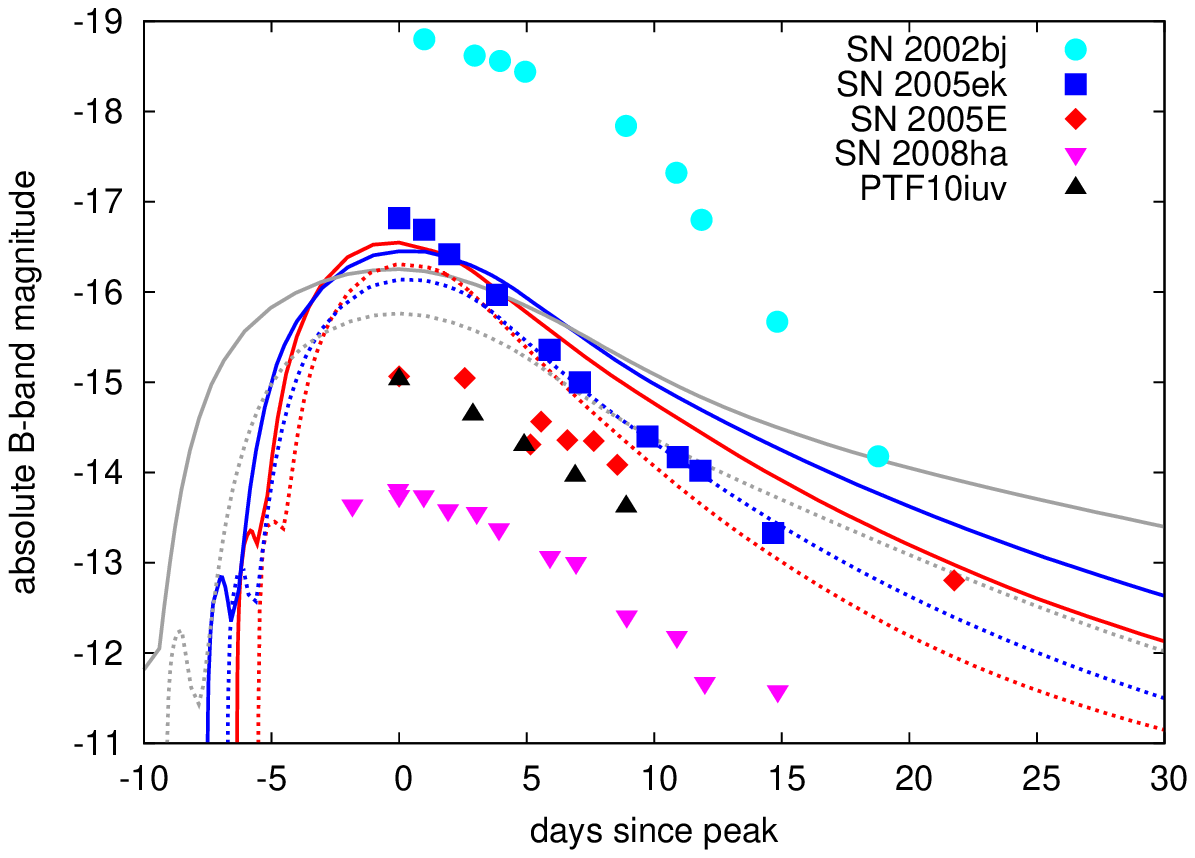} 
 \end{center}
\caption{
Comparison between our synthetic and observed bolometric (top) and $B$-band 
(bottom) LCs of rapidly-evolving SNe. These are from \citet{drout2014} 
(PS1-12bb), \citet{kasliwal2010} (SN~2010X), \citet{drout2013} (SN~2005ek), 
\citet{poznanski2010} (SN~2002bj), \citet{kasliwal2012} (PTF10iuv), 
\citet{perets2010} (SN~2005E), and \citet{foley2009} (SN~2008ha).
}\label{fig:lcobs}
\end{figure}

\begin{figure*}
 \begin{center}
  \includegraphics[width=\columnwidth]{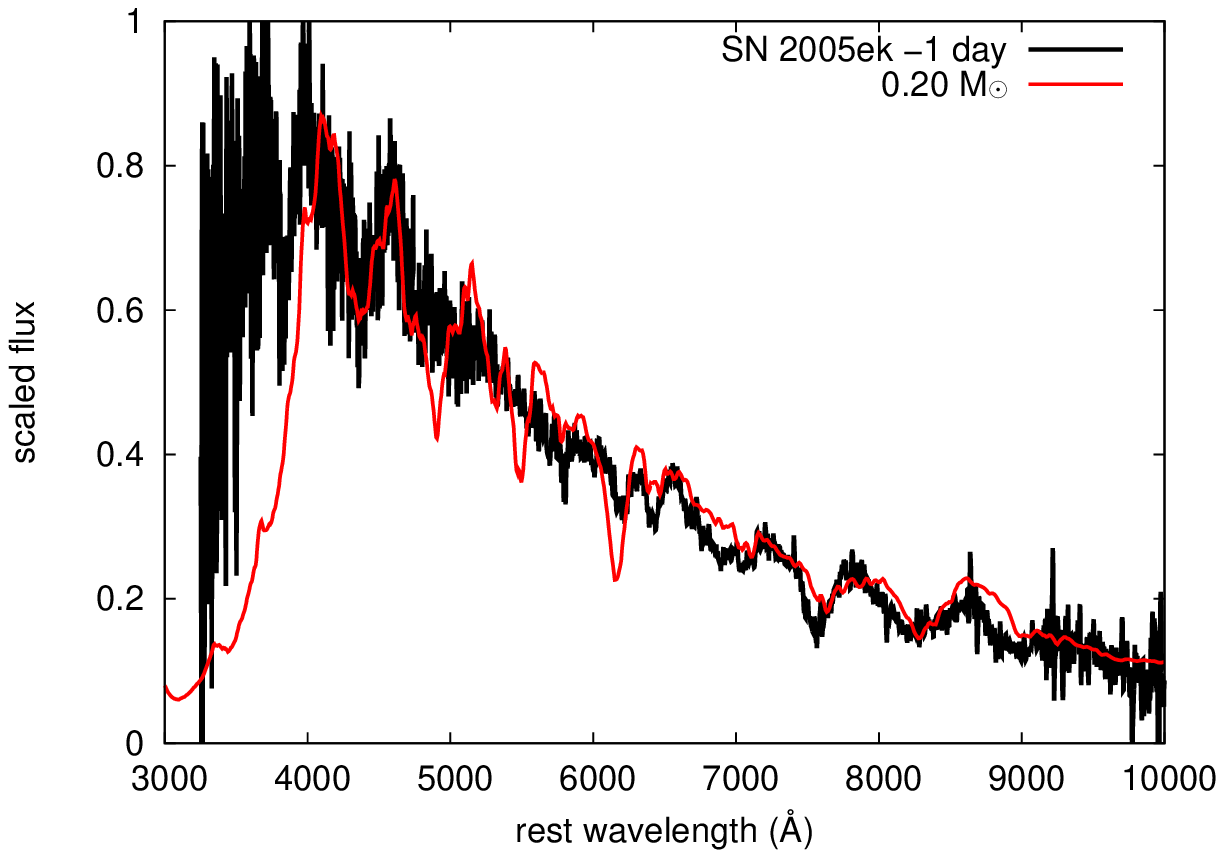} 
  \includegraphics[width=\columnwidth]{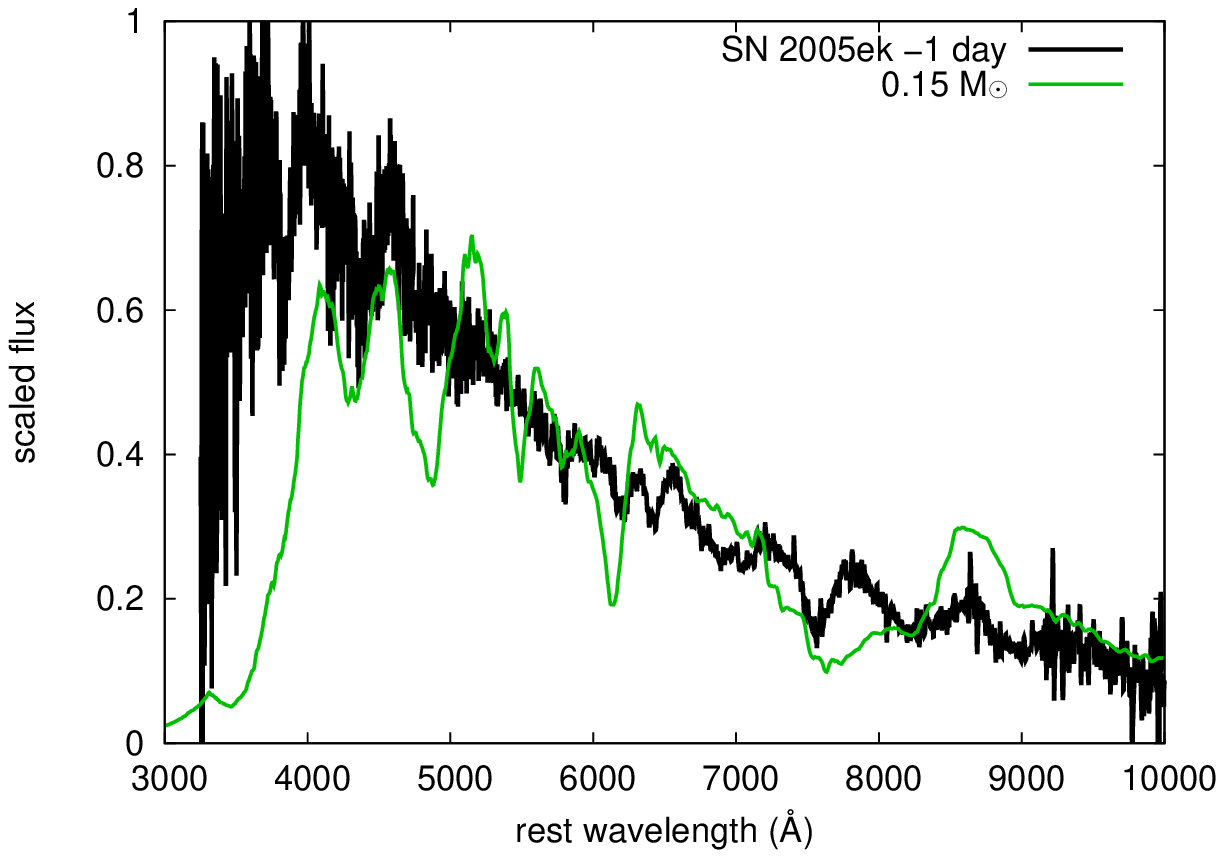} 
 \end{center}
\caption{
Comparison between our synthetic spectra and the observed spectrum of SN~2005ek 
one day before LC peak from \citet{drout2013}.
}\label{fig:spsn2005ek}
\end{figure*}

\begin{figure}
 \begin{center}
  \includegraphics[width=\columnwidth]{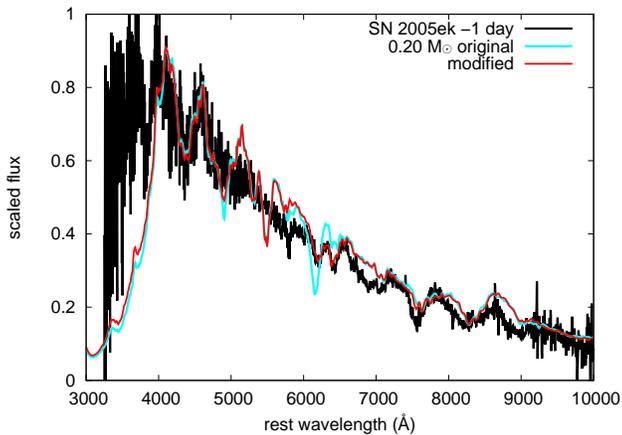} 
 \end{center}
\caption{
Synthetic spectrum for the $\Mej=0.20~\Msun$ and $\Eej=0.25~\mathrm{B}$ model 
where the Si abundance is artificially reduced and the O abundance is 
artificially increased. The model provides a better match than the original one.
}\label{fig:spsn2005ekmod}
\end{figure}

\section{Comparison with observations}\label{sec:compobs}
\subsection{Light curves}\label{sec:lccomp}
Figure~\ref{fig:lcobs} shows a comparison between our synthetic LCs and
observational data of rapidly-evolving SNe. The comparison is shown both in
bolometric luminosity and in the $B$~band. Overall, our LCs are consistent with faint
rapidly-evolving SNe with peak luminosity of 
$\sim 10^{42}~\mathrm{erg~s^{-1}}$ (i.e., $\sim -16$~mag). 
Luminous SNe like SN~2002bj require more than 0.1~\Msun\ of \Ni\ to explain the 
peak luminosity by \Ni, which is inconsistent with the small amount of \Ni\ 
($\sim 0.01~\Msun$) we expect in our ultra-stripped SNe. Ultra-stripped SN LCs 
are consistent with those of rapidly-evolving faint SNe (e.g., SN~2010X, 
PS1-12bb, and SN~2008ha), the so-called Ca-rich gap transients 
(e.g., PTF10iuv and SN~2005E), and SN~2005ek.

\subsection{Spectra}\label{sec:spcomp}

We compare our models with observed near-LC-peak spectra of some rapidly-evolving SNe , i.e.,
SN~2005ek, two Ca-rich gap transients (PTF10iuv and SN~2005E), SN~2010X, SN~2002bj, and
a SN~2002cx-like SN 2007qd.
The observed spectra are taken from
WISeREP\footnote{\url{http://wiserep.weizmann.ac.il}} \citep{yaron2012}.
We use synthetic spectra
of 0.25~B for comparison, as there are no significant differences in spectral
features (except for velocity) caused by the difference in the explosion energy
in most spectra (Fig.~\ref{fig:peaksp}).
We correct the extinction of observed spectra by using the Galactic extinction law
of \citet{cardelli1989} assuming $R_V = 3.1$.
The redshifts and extinctions applied are,
$z=0.017$ and $E(B-V)=0.21~\mathrm{mag}$ for SN~2005ek \citep{drout2011},
$z=0.02$ and $E(B-V)=0~\mathrm{mag}$ for PTF10iuv \citep{kasliwal2012},
$z=0.0090$ and $E(B-V)=0.098~\mathrm{mag}$ for SN~2005E \citep{perets2010},
$z=0.015$ and $E(B-V)=0.146~\mathrm{mag}$ for SN~2010X \citep{kasliwal2010},
$z=0.012$ and $E(B-V)=0~\mathrm{mag}$ for SN~2002bj \citep{poznanski2010}, and
$z=0.043$ and $E(B-V)=0.035~\mathrm{mag}$ for SN~2007qd \citep{mcclelland2010}.

\subsubsection{SN 2005ek}

SN~2005ek was suggested to be a SN whose progenitor experienced extensive  mass
stripping by binary interactions \citep{drout2013}. Its LC was shown to be
consistent with our ultra-stripped SN model \citep{tauris2013}, but the spectral
properties have not yet been compared.

Figure~\ref{fig:spsn2005ek} shows a comparison between synthetic spectra
obtained from our ultra-stripped SN models and the spectrum of SN~2005ek one day
before LC peak. Overall, continuum features and velocity shifts of
the synthetic spectra match the observed spectrum well. In particular, the
characteristic features in the red part of the spectrum of SN~2005ek, i.e.,
C~\textsc{i} $\lambda 6582$, O~\textsc{i} $\lambda 7774$, and Ca~\textsc{ii} IR
triplet, are all well reproduced by the $\Mej=0.20~\Msun$ model.

Our models predict stronger Si~\textsc{ii} features than seen in SN~2005ek. We
find that we can obtain a better match by reducing the Si abundance from 0.053
to 0.003 and increasing the O abundance from 0.18 to 0.23
(Fig.~\ref{fig:spsn2005ekmod}). The strong Si features in the original
ultra-stripped SN model may also be due to our assumption of fully-mixed ejecta.
The average Si abundance used in the spectral modelling is around 0.05, while
most of the outer layers have smaller fractions initially (Fig.~\ref{fig:nuc}).
The smaller abundances may result in weaker Si features especially at early
times. Thus, our assumption of full mixing may be too simplistic and SN~2005ek
might have had a rather smaller degree of mixing. This may also explain the
significant drop in flux below $\sim 3500$\,\AA\ in the model: a larger degree
of mixing results in more Fe-group elements in the outer layers, leading to more
line blocking which suppresses the NUV flux.

\begin{figure*}
 \begin{center}
  \includegraphics[width=\columnwidth]{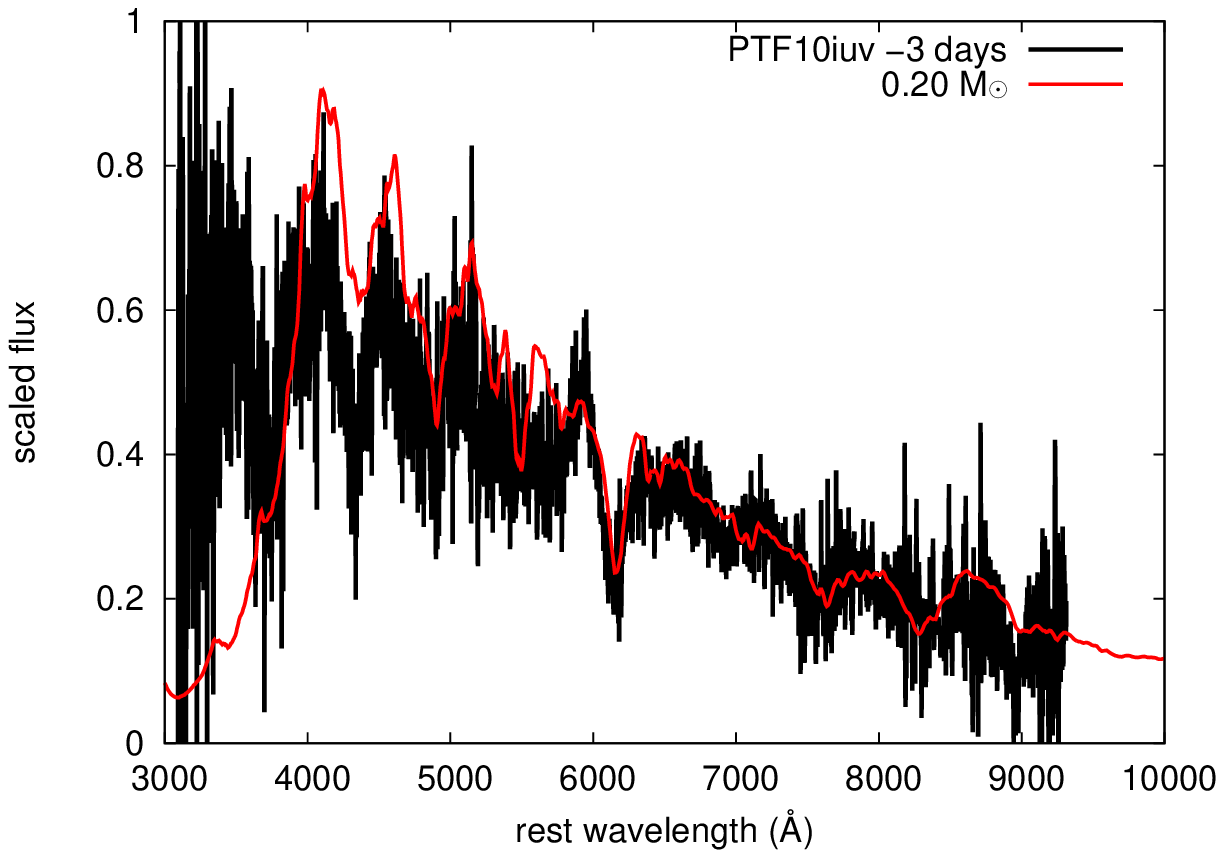} 
  \includegraphics[width=\columnwidth]{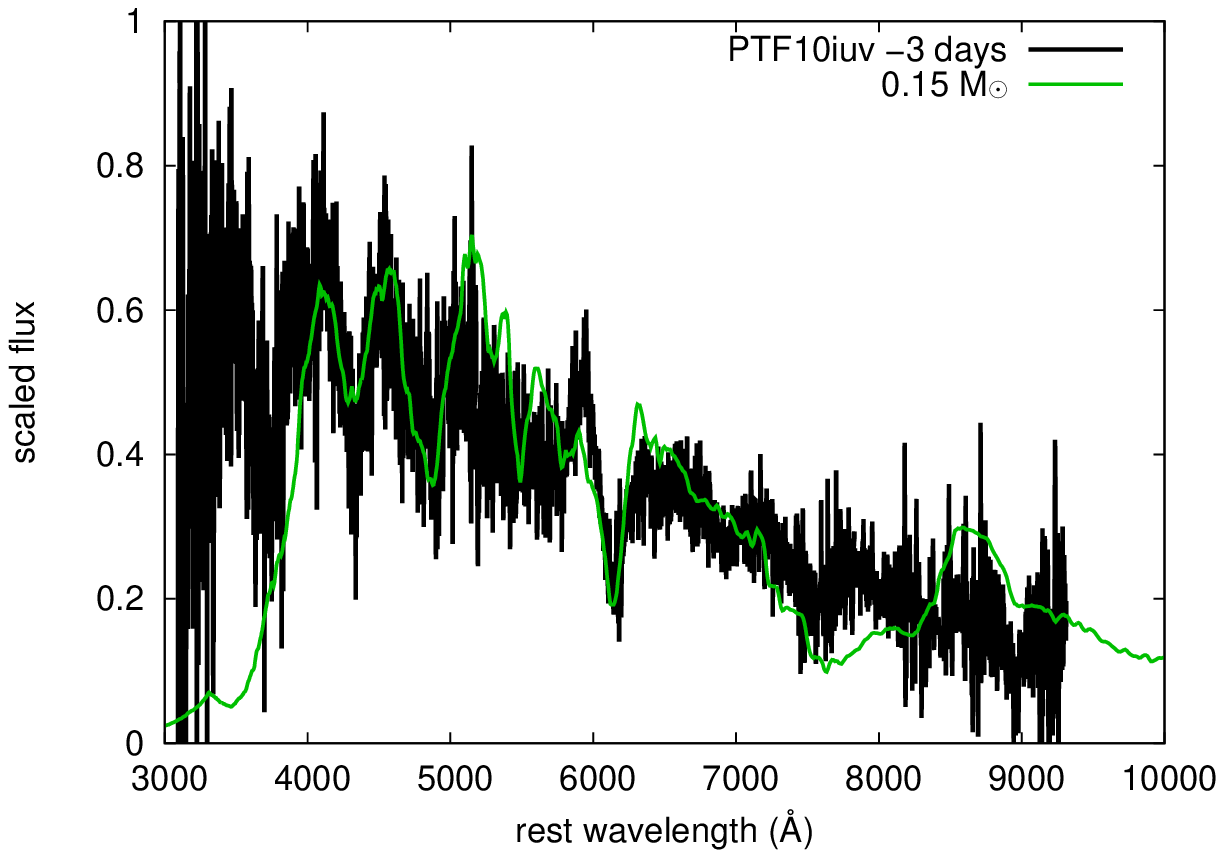} \\
  \includegraphics[width=\columnwidth]{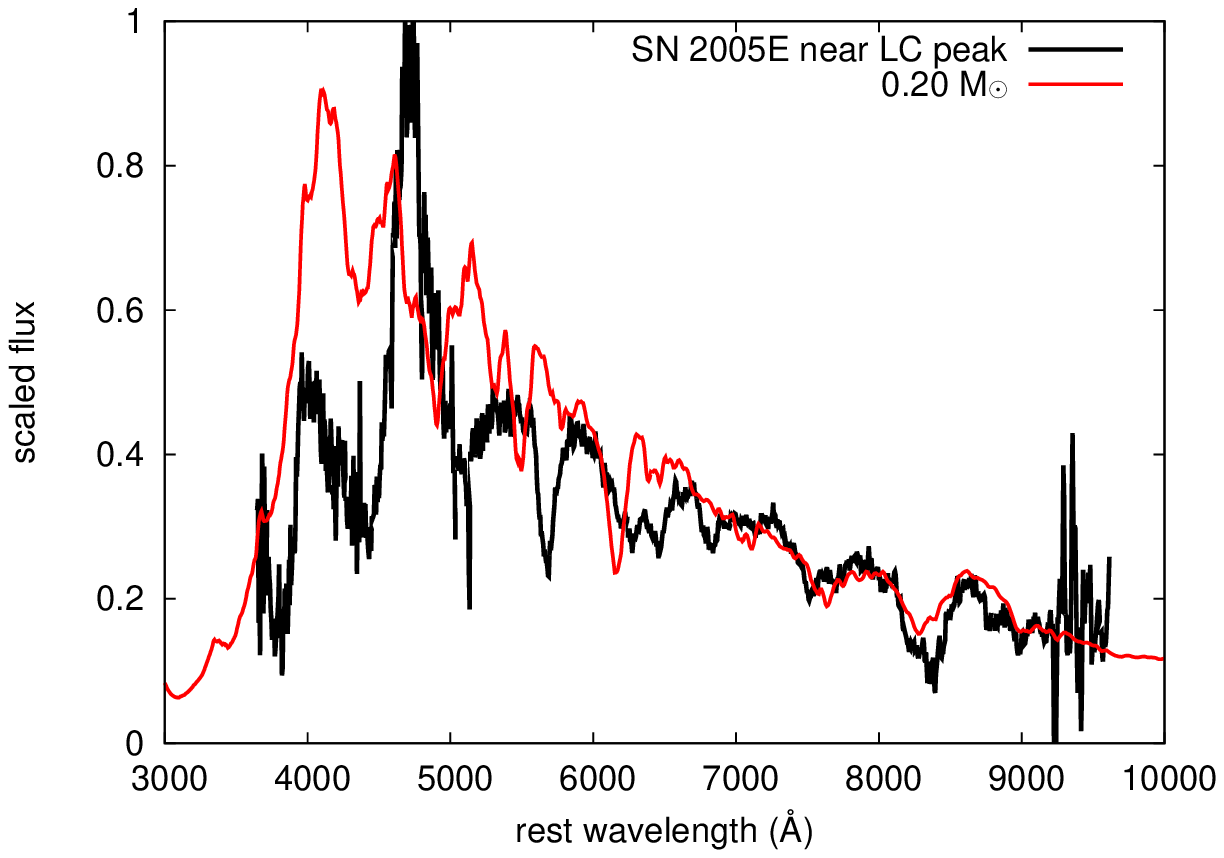} 
  \includegraphics[width=\columnwidth]{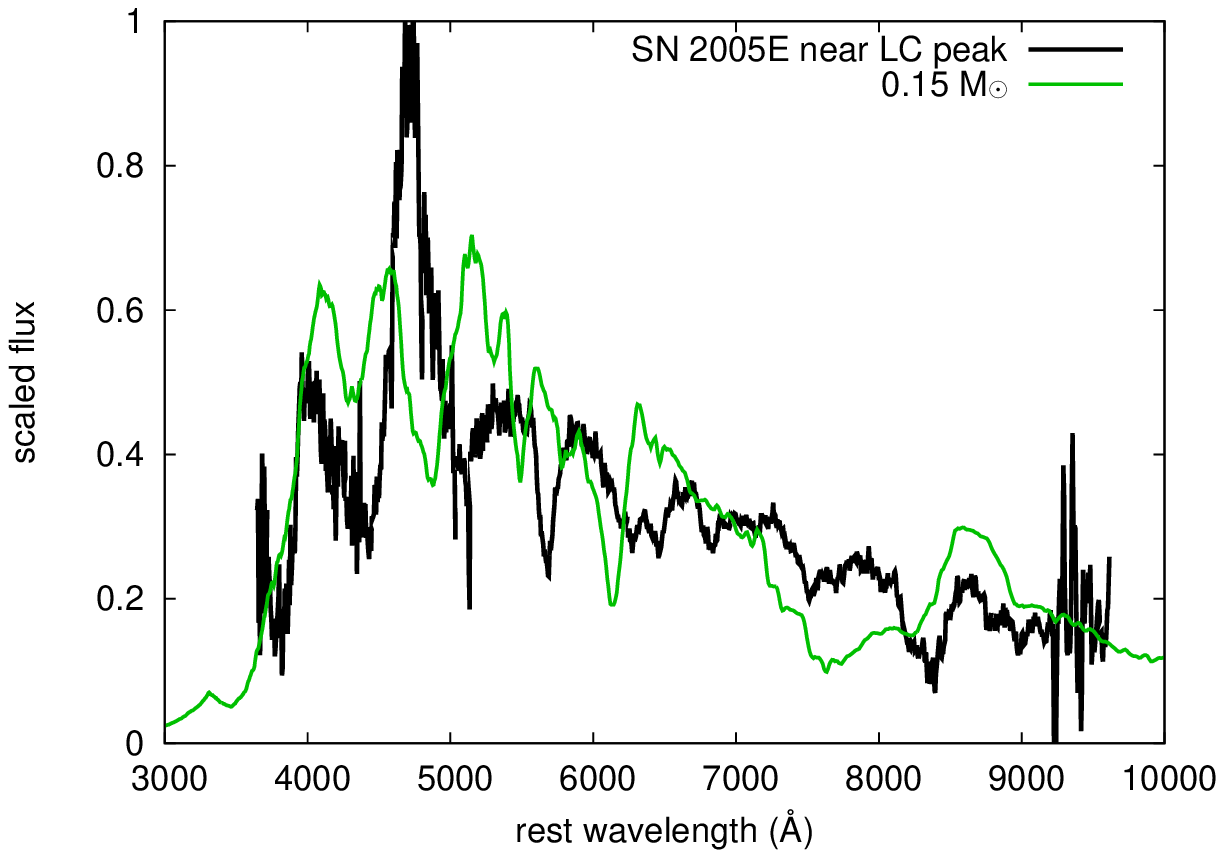} 
 \end{center}
\caption{
Comparison between our synthetic spectra and the observed near-maximum spectra 
of the Ca-rich gap transients PTF10iuv \citep{kasliwal2012} and SN~2005E 
\citep{perets2010}.
}\label{fig:spcarich}
\end{figure*}

\subsubsection{Ca-rich gap transients (PTF10iuv and SN~2005E)}

Fig.~\ref{fig:lcobs} shows the LCs of two
Ca-rich gap transients, PTF10iuv and SN~2005E. Their peak luminosity is slightly
smaller than that of our ultra-stripped SNe, but the timescales of the LC
evolution as well as the LC decline rates are similar. Efficient mixing in the
ejecta may make the luminosity of our models slightly lower
(Fig.~\ref{fig:lcmix}), which motivates us to investigate their spectra. 

Figure~\ref{fig:spcarich} shows a spectral comparison near peak luminosity. The
spectral features including velocity shifts of PTF10iuv (top panels of
Fig.~\ref{fig:spcarich}) match our models well. Prominent features in PTF10iuv,
such as Si~\textsc{ii} $\lambda 6347$, O~\textsc{i} $\lambda 7772$, and
Ca~\textsc{ii} $\lambda 8542$ are also well reproduced. We find relatively
strong Si and Ca lines from our core-collapse progenitors because of the small
values of the explosion energy and the progenitor mass. The overall color of
PTF10iuv is consistent with our model. In summary, the LC and spectral properties of
PTF10iuv are overall consistent with our ultra-stripped SN models.

On the other hand, our ultra-stripped SN spectral models are not consistent with the spectral
features of the other Ca-rich gap transient, SN~2005E, very well, as shown in
the bottom panels of Fig.~\ref{fig:spcarich}. This indicates that there may be
several kinds of progenitors for Ca-rich gap transients, of which some could
be related to ultra-stripped SNe.

Ca-rich gap transients are typically found in remote locations from the center
of their host galaxies \citep{kasliwal2012,lyman2014b,lyman2016,foley2015}. For
example, PTF10iuv was 40~kpc away from the closest host galaxy candidate
\citep{kasliwal2012}. This fact is used to relate Ca-rich gap transients to
explosive events related to white dwarfs. However, Ca-rich transients are also
found in the intergalactic space of merging galaxies \citep[e.g.,][]{foley2015},
where star formation is likely to take place \citep[e.g.,][]{mullan2011}.
Some massive stars are also known to exist very far from
apparent star forming regions \citep[e.g.,][]{smith2016}.
It is also interesting to note that white dwarf mergers may actually end up as 
core-collapse SNe and could lead to SNe with similar properties to our
ultra-stripped SNe \citep[e.g.,][]{nomoto1985,schwab2016}.
As we suggest here, Ca-rich gap
transients can have different origin. Those found far from any host galaxy may
be unrelated to ultra-stripped SNe. The faint nature of ultra-stripped SNe may
also prevent them from being found in bright star-forming regions such as in
galactic disks.

\begin{figure*}
 \begin{center}
  \includegraphics[width=\columnwidth]{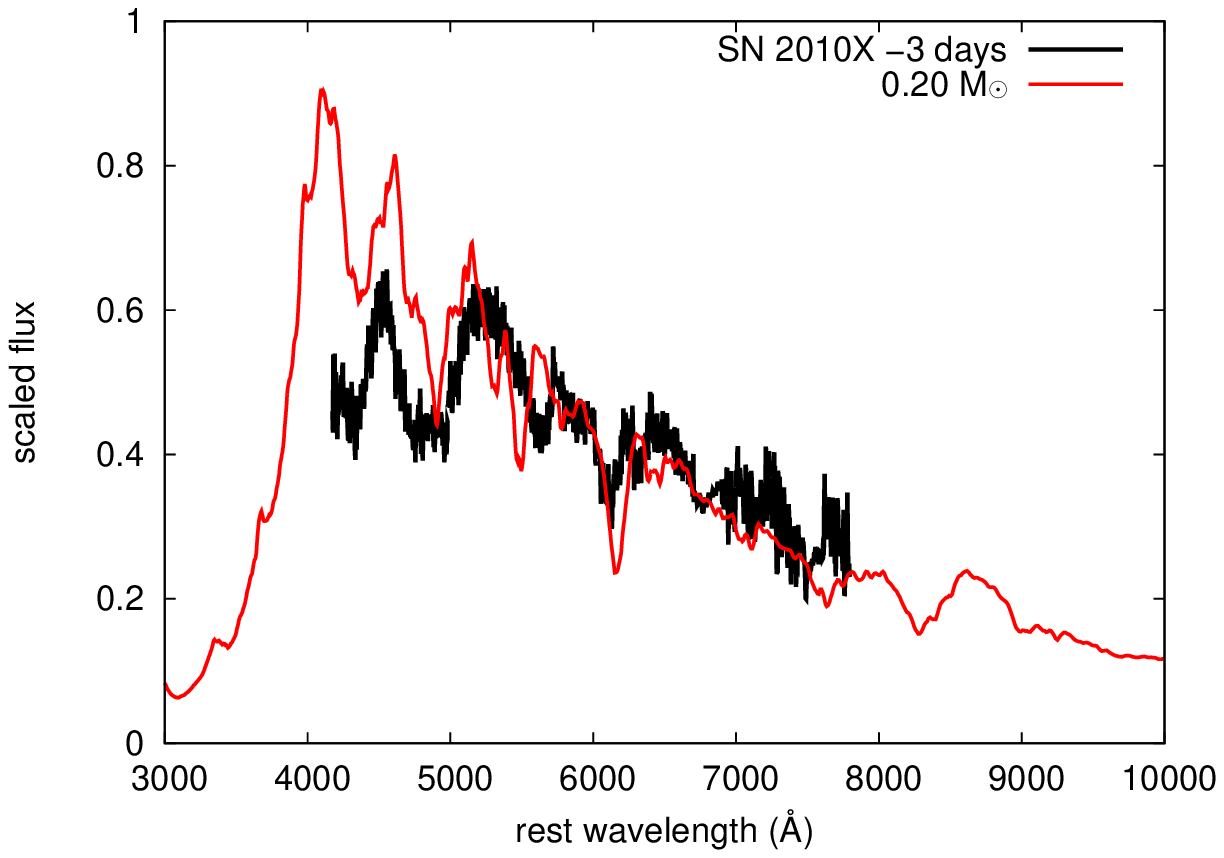}
  \includegraphics[width=\columnwidth]{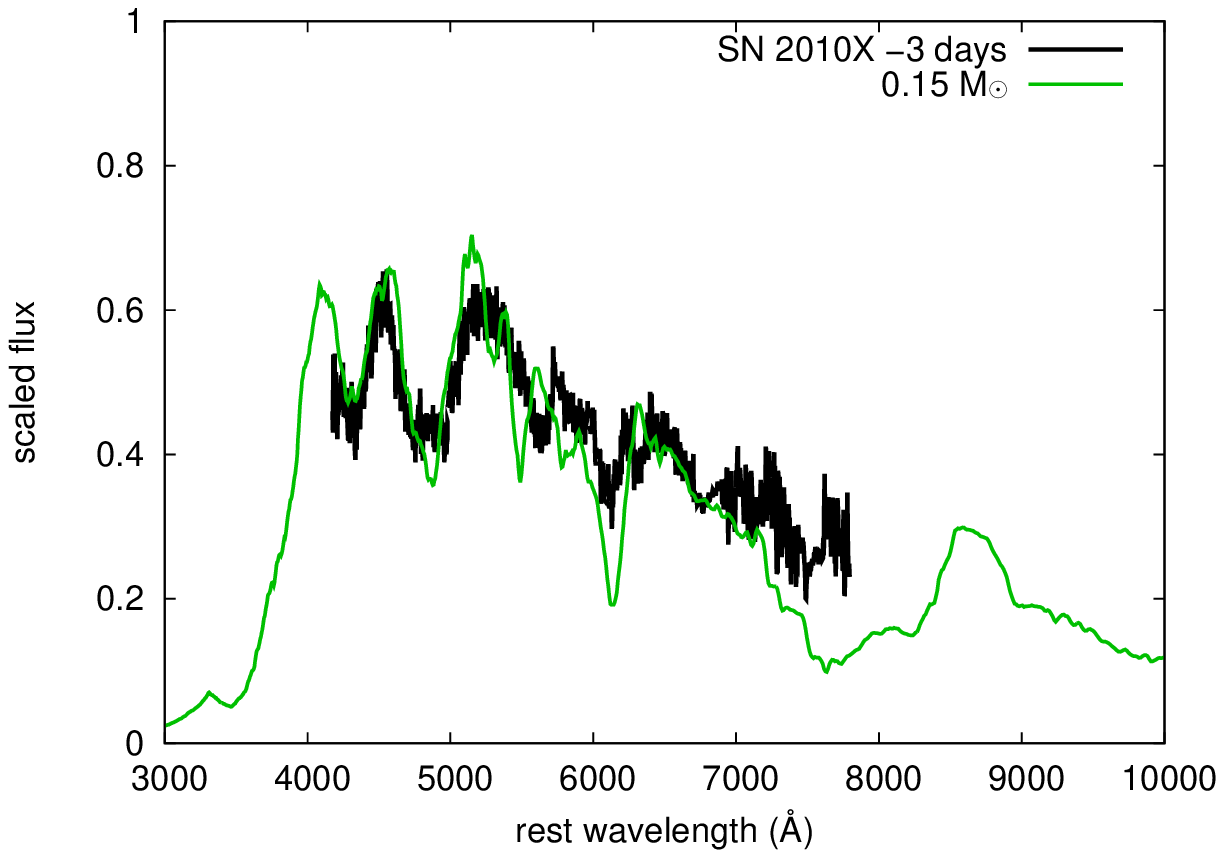} 
 \end{center}
\caption{
Comparison between our synthetic spectra and the spectrum of SN~2010X 
\citep{kasliwal2010} near LC peak.
}\label{fig:spsn2010X}
\end{figure*}

\begin{figure}
 \begin{center}
 \includegraphics[width=\columnwidth]{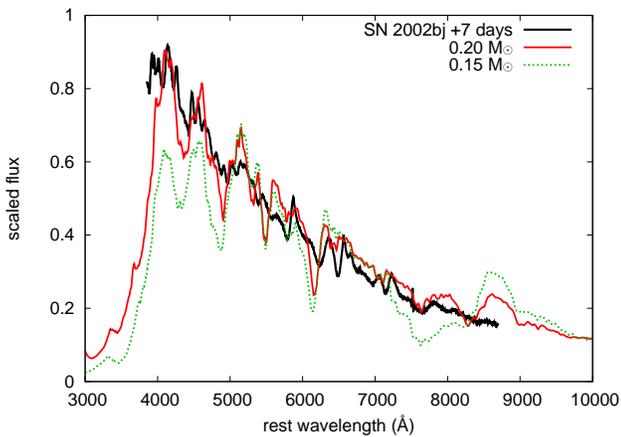}
 \end{center}
\caption{
Comparison between our synthetic spectra at maximum and the earliest observed 
spectrum of SN~2002bj \citep{poznanski2010}.
}\label{fig:spsn2002bj}
\end{figure}

\begin{figure}
 \begin{center}
  \includegraphics[width=\columnwidth]{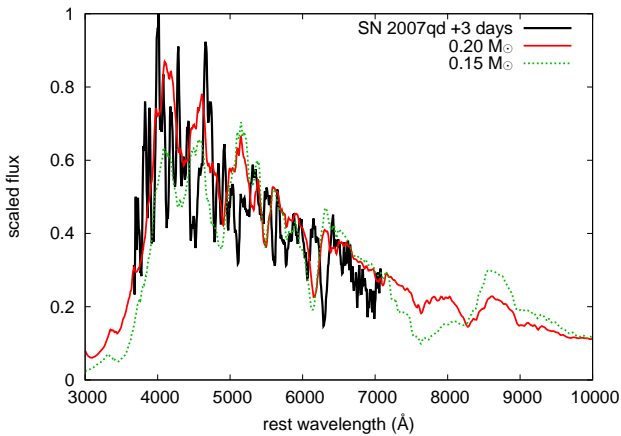}
 \end{center}
\caption{
Comparison between our synthetic spectra at maximum and the observed spectrum of 
the SN~2002cx-like SN~2007qd near maximum \citep{mcclelland2010}.
}\label{fig:spsn2002cxlike}
\end{figure}

\subsubsection{SN 2010X}

SN~2010X was a rapidly-evolving Type~Ib SN. Its ejecta mass was estimated to be
$\sim 0.16$~\Msun\ and its \Ni\ mass to be $\sim 0.02$~\Msun\ \citep{kasliwal2010}.
These properties, as well as the LC evolution are similar to those of our
ultra-stripped SNe (Fig.~\ref{fig:lcobs}). SN~2010X had spectral characteristics
similar to SN~2002bj, which we discuss in the next section, but it was much
fainter. 

We compare our spectra with that of SN~2010X near LC peak in
Fig.~\ref{fig:spsn2010X}.
Overall spectral features match well, especially the model
with $\Mej=0.15~\Msun$. The Si~\textsc{ii} $\lambda 6355$ is predicted to be
slightly stronger, but a small reduction of the Si abundance or a slightly
smaller degree of mixing could reduce the strength of this line as was discussed
for SN~2005ek. Therefore, SN~2010X may also be related to ultra-stripped SNe.

\subsubsection{SN 2002bj}

SN~2002bj was one of the first rapidly-evolving SNe to be reported
\citep{poznanski2010}. If SN~2002bj was powered by \Ni, the amount of \Ni\
required to explain its peak luminosity is $0.15-0.25~\Msun$
(\citealt{poznanski2010}, see also Fig.~\ref{fig:lcobs}). The amount of \Ni\
required is much larger than what we expect from ultra-stripped SNe, and the
\Ni\ mass alone is comparable to the total ejecta mass expected from
ultra-stripped SNe. This suggests that SN~2002bj is unlikely to be related to
ultra-stripped SNe. Nonetheless, we show the comparison between our synthetic
spectra and that of SN~2002bj.
As expected, the match is not very good (Fig.~\ref{fig:spsn2002bj}).

\subsubsection{SN 2002cx-like (Type~Iax) SNe}

SN~2002cx-like SNe, which are often referred as Type~Iax SNe, are a peculiar
type of SN~Ia with fainter peak luminosity. Their origin is still under
discussion \citep[e.g.,][]{foley2016}. Although many of them have peak
magnitudes brighter than $-17$~mag (see \citealt{foley2016} for a summary) and
are thus too bright to be ultra-stripped SNe, some of them do have fainter peak
luminosity \citep[e.g.,][]{stritzinger2014,mcclelland2010}. We take one faint
SN~2002cx-like SN, SN~2007qd \citep{mcclelland2010}, which is in the expected
peak luminosity range for ultra-stripped SNe, and compare its spectrum at the
peak luminosity with our synthetic spectra.

Figure~\ref{fig:spsn2002cxlike} shows the comparison. SN~2002cx-like SNe have
much lower velocities than our synthetic ultra-stripped SN spectra. A low
photospheric velocity is a commonly observed feature of faint SN~2002cx-like
SNe. Our model with the smallest explosion energy (0.10~B) may have a
photospheric velocity similar to that of some SN~2002cx-like SNe
(Fig.~\ref{fig:photo}), but it still seems to fail to reproduce crowded,
unblended line features. In addition, the spectra of ultra-stripped SNe evolve
quickly to the nebular regime, while the spectra of SN~2002cx-like SNe do not
\citep[e.g.,][]{sahu2008,foley2016}.

\subsection{Summary}

Table~\ref{table:stars} summarizes the results of the comparison between our
ultra-stripped SN models and possible observational counterparts.
We find that SN~2005ek, PTF10iuv (Ca-rich gap transient), and SN~2010X match our
synthetic ultra-stripped SN LCs and spectra.

\begin{table}
\centering
\caption{Ultra-stripped SN candidates. An `\checkmark' or an `x' indicate whether or 
not the SN properties match those expected of ultra-stripped SNe.}
\label{table:stars}
\begin{tabular}{lcccc}
\hline
SN  & LC & velocity & Si & Ca  \\
\hline
SN~2005ek& \checkmark & \checkmark & \checkmark & \checkmark \\
PTF10iuv (Ca-rich gap)& \checkmark & \checkmark & \checkmark & \checkmark \\
SN~2005E (Ca-rich gap)& \checkmark & x & x & \checkmark \\
SN~2010X & \checkmark & \checkmark & \checkmark & \checkmark \\
SN~2002bj & x & x & x & x \\
SN~2007qd (02cx-like) & \checkmark & x & \checkmark & ? \\
\hline
\end{tabular}
\end{table}

\section{Discussion}\label{sec:discussion}
\subsection{Diversity}

We have investigated the observational properties of an ultra-stripped SN
explosion from a  progenitor with a mass of 1.50~\Msun. Depending on the initial
binary parameters and the helium ZAMS mass of the progenitor, the final total
mass and core mass vary considerably \citep{tauris2015}. However, recent
numerical simulations of explosions of ultra-stripped core-collapse SNe find
that the explosion energy and the synthesized \Ni\ mass do not change
significantly among progenitors of different core masses \citep{suwa2015}. Thus,
the explosion energy and \Ni\ mass values investigated in this paper are likely
to remain similar even in different ultra-stripped SN progenitors.
The low spread in the \Ni\ masses found in our models
suggests that the peak luminosity should be roughly similar in all
ultra-stripped SNe, although the LC rise time can vary depending on the ejecta
mass \citep{tauris2015}. In our one-dimensional model the \Ni\ mass ejected
depends on the mass cut, but \citet{suwa2015} obtained similar \Ni\ masses in
their multi-dimensional explosion simulations. Thus, the spectral signatures we
find in this study for a particular progenitor model are likely to be generic
for ultra-stripped SNe.

A possible major source of diversity in ultra-stripped SN properties
that may be caused by
differences in their progenitors is their SN spectral type, which is determined
by the ejected helium mass. In the progenitor model we used in this study, there
are 0.03~\Msun\ of helium in an ejected mass of $0.15-0.20~\Msun$. Although
helium features are not expected to be observed significantly in our model and
we expect the explosion to be a SN~Ic  (Section~\ref{sec:hefeatures}), the
helium mass as well as the ejecta mass can change depending on the initial
binary configurations \citep{tauris2015}. Many progenitors are expected to have
a helium mass above the critical helium mass ($\sim 0.1~\Msun$,
\citealt{hachinger2012}) required to observe optical helium features
\citep{tauris2015}, and thus may be observed as SNe~Ib like SN~2010X.

\begin{figure}
 \begin{center}
  \includegraphics[width=\columnwidth]{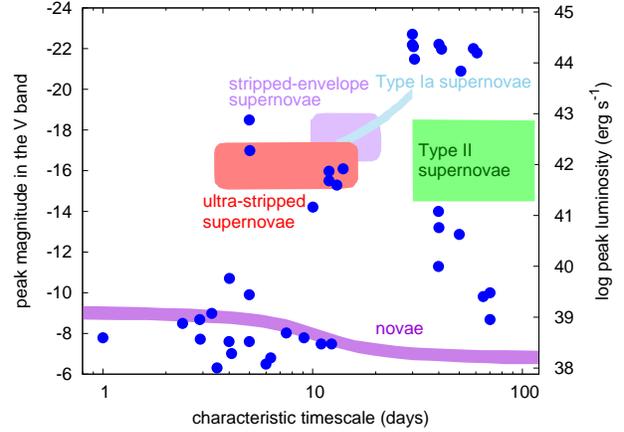}
 \end{center}
\caption{
Characteristics of ultra-stripped SNe in the phase-space diagram of optical 
transients from \citet{kulkarni2012}.
}\label{fig:kulkarni}
\end{figure}

To summarize, the expected luminosity range of ultra-stripped SNe is similar to
that obtained with the model used in this paper. The diversity in SN ejecta
masses caused by different progenitor systems could lead to diversity in the
rise times of ultra-stripped SNe. We indicate the expected location of
ultra-stripped SNe in the phase diagram of transients
\citep[e.g.,][]{kulkarni2012} in Fig.~\ref{fig:kulkarni}. 
It is likely that there is a smooth transition between ultra-stripped SNe
and stripped-envelope SNe. The classical Type~Ic SN~1994I, which
had an ejecta mass of only 1~\Msun \citep[e.g.,][]{iwamoto1994,sauer2006} may be
an example of a SN~Ic located between typical stripped-envelope SNe and
ultra-stripped SNe. Also, the low-mass Type~Ic SN~2007gr
\citep{hunter09,mazzali2010} may be an even more extreme case.

\subsection{Event rates}

Based on our spectral analysis, we have shown that SN~2005ek and some of the
Ca-rich gap transients, as well as SN~2010X, might be related to ultra-stripped
SNe. The event rate of rapidly-evolving SNe including SN~2005ek, SN~2010X, and
SN~2002bj are estimated to be at least 1-3 per cent of SNe~Ia \citep{drout2013}.
The total event rate of the Ca-rich gap transients is estimated to be at least a
few per cent of SNe~Ia \citep{kasliwal2012,perets2010}. Given that a part of the
transients included to estimate the two event rates actually correspond to
ultra-stripped SNe, we can roughly estimate that the event rate for the
ultra-stripped SNe based on SN observations is at least a few per cent of
SNe~Ia. Because the volumetric rates of SNe~Ia and SNe~Ib/Ic are similar (e.g.,
\citealt{li2011}), ultra-stripped SN rates are presumed to be at least a few per
cent of SNe~Ib/Ic. This rate corresponds to roughly 1 per cent of all
core-collapse SNe including SNe~II \citep{li2011} and matches the rough event
rate estimate by \citet{tauris2013} ($0.1-1$ per cent of core-collapse SNe).
Further observational and theoretical studies to better estimate the event rate
of ultra-stripped SNe are encouraged for their better comparison.

\subsection{Synergy with gravitational-wave astronomy}

Because ultra-stripped SNe are presumed to lead to double-NS systems, but not
all double NS systems merge within a Hubble time, the observed rate of
ultra-stripped SNe is expected to be larger than the rate of double-NS mergers.
Furthermore, ultra-stripped SNe may also originate in binary systems with a
white dwarf or a black hole accretor \citep{tauris2013}. After the recent
success in detecting gravitational wave signals from merging compact objects by
advanced LIGO \citep{abbott2016}, it is expected that gravitational waves from
merging double-NS systems will be detected in the near future, with advanced
LIGO or other detectors like advanced VIRGO and KAGRA. Thus, it will soon be
possible to obtain NS merger rate constraints from gravitational wave
observations. This should place a lower limit to the rate of ultra-stripped SNe.
Comparing the rates of ultra-stripped SNe and NS mergers, we will be able to test
whether the ultra-stripped SN channel is actually the major path to form
double-NS systems, and thus to constrain binary stellar evolution using
gravitational waves. 

The current constraints on the NS merger rate from LIGO is
less than $12600~\mathrm{Gpc^{-3}~yr^{-1}}$ \citep{abbott2016b}.
Assuming the Galactic core-collapse SN rate of $\sim 0.01~\mathrm{yr^{-1}}$ and
ultra-stripped SN fraction of $0.1-1$ per cent in core-collapse SNe, we obtain
an expected Galactic ultra-stripped SN rate of $\sim 10^{-5}-10^{-4}~\mathrm{yr^{-1}}$.
If we adopt the Milky Way equivalent galaxy density of $0.01~\mathrm{Mpc^{-3}}$ \citep{kopparapu2008},
we expect an ultra-stripped SN rate of $\sim 100-1000~\mathrm{Gpc^{-3}~yr^{-1}}$.
This is significantly lower than the current upper limit on the NS-NS merger rate
from LIGO.

Finally, it should be noted that the distribution of eccentricities in current observations of 
 NS-NS binaries \citep{martinez2015,lazarus2016} supports the idea that ultra-stripped SNe are indeed the 
 progenitors of the second SN in these systems, and that they often (but not always) result in small kicks. 
 Out of 12 NS-NS systems, 9 have an eccentricity of less than 0.3. Moreover, even relatively small NS kicks of $50\;{\rm km\,s}^{-1}$ 
 can result in post-SN eccentricities of more than $0.5$, depending on the orbital period (Tauris et al. in preparation).

\subsection{Observations of ultra-stripped SN progenitors}
While it has been argued that the descendants of ultra-stripped SNe are often close pairs of NSs in binaries,
it is much more difficult to find direct observational evidence for the progenitors of ultra-stripped SNe.
Arguably, a fraction of high-mass X-ray binaries currently containing a NS and an OB-star will eventually evolve into double NS systems.
Others will merge or become disrupted in the process. However, in between these two stages, the immediate
progenitors of ultra-stripped stars would be found either as post-common envelope naked helium stars (Wolf-Rayet stars)
or in an X-ray binary during the subsequent so-called Case~BB Roche-lobe overflow. However, these phases are
short lasting (especially the latter which is typically $<10^5\;{\rm yr}$, \citealt{tauris2015}) and thus chances
of detecting these systems are small, even though the helium star is rather luminous ($>10^4\;L_{\odot}$). One possible related observed system is Cygnus X-3 which is a close binary system with a NS or BH and a Wolf-Rayet star, but the mass of the Wolf-Rayet star (about 10~\Msun, \citealt{zdziarski2013}) is too high to correspond to an ultra-stripped SN progenitor.

\section{Conclusions}\label{sec:conclusions}

We have presented synthetic LCs and spectral properties of ultra-stripped SNe.
We evolved the ultra-stripped SN progenitor presented previously
\citep{tauris2013} until core collapse, calculated its explosive
nucleosynthesis, and then synthesized LCs and spectra. Our ultra-stripped SNe
have explosion energies of $1-5\times 10^{50}~\mathrm{erg}$ and ejecta masses of
$\sim 0.1$\,\Msun. Explosive nucleosynthesis calculations show that they produce
about $0.03~\Msun$ of \Ni. We also found that ultra-stripped SNe have rise times
of $5-10$~days and their peak luminosity is $\sim -16$~mag or
$10^{42}~\mathrm{erg~s^{-1}}$.

Several types of transients have been found that have rise times and peak
luminosities similar to those expected for ultra-stripped SNe. They show diverse
spectral properties. We compared our synthetic spectra with those of
rapidly-evolving transients showing LCs similar to those of our synthetic
ultra-stripped SN LCs, and found that the spectra of SN\,2005ek, some of
so-called Ca-rich gap transients, and SN\,2010X match reasonably well our
synthetic ultra-stripped SN spectra. Not all Ca-rich gap transients have similar
properties to our ultra-stripped SNe, indicating that this group of transients
may include events with different origin. For example, the spectra of PTF10iuv
are consistent with our ultra-stripped SN spectra, while those of SN~2005E are
not. If all the transients above mentioned are actually from ultra-stripped SNe,
the event rate of ultra-stripped SNe would be about one per cent of all
stripped-envelope SNe.

It has been suggested that ultra-stripped SNe may be a major evolutionary path
to form double-NS systems which could merge within a Hubble time and that
double-NS systems left by ultra-stripped SNe may dominate the population of
merging double-NS systems which is expected to be observed by GW observatories
\citep{tauris2013,tauris2015}. If this is true, we expect the NS merger rate to
be comparable to or somewhat smaller than that of ultra-stripped SNe.

\section*{Acknowledgments}
TJM thanks Yudai Suwa and Markus Kromer for helpful discussions.
TJM is supported by Japan Society for the Promotion of Science Postdoctoral Fellowships for Research Abroad (26\textperiodcentered 51) and by the Grant-in-Aid for Research Activity Start-up of the Japan Society for the Promotion of Science (16H07413).
The work of S.Blinnikov on development of STELLA code is supported by
Russian Science Foundation grant 14-12-00203.
The work has also been supported by a Humboldt Research Award to PhP at
the University of Bonn.
Numerical computations were partially carried out on Cray XC30 and PC cluster at Center for Computational Astrophysics, National Astronomical Observatory of Japan.
The numerical calculations were also partly carried out on Cray XC40 at Yukawa Institute for Theoretical Physics in Kyoto University.
We made use of the Weizmann interactive supernova data repository - \url{http://wiserep.weizmann.ac.il}.
This research has made use of the NASA/IPAC Extragalactic Database (NED) which is operated by the Jet Propulsion Laboratory, California Institute of Technology, under contract with the National Aeronautics and Space Administration.

\label{lastpage}

\end{document}